%% file: PCA_paper_v9.tex
\documentclass[numberedappendix,appendixfloats]{emulateapj}
\usepackage{graphicx}
\usepackage{epsfig}
\usepackage{amsmath}
\usepackage{amssymb}
\usepackage{multirow}
\usepackage{natbib}
\usepackage{float}
\usepackage{verbatim}
\usepackage{color}
\usepackage{subfigure}

\renewcommand{\t}{\text}

\begin{document}
\title{What Do Dark Matter Halo Properties Tell Us About Their Mass Assembly Histories?}
\shorttitle{What Do Halo Properties Tell Us About Mass Assembly?}
\shortauthors{Wong \& Taylor}

\author{Anson W.C. Wong\altaffilmark{1} and James E. Taylor\altaffilmark{1}}           
\submitted{Submitted to ApJ}
\email{a57wong@uwaterloo.ca, taylor@uwaterloo.ca}

\altaffiltext{1}{Department of Physics and Astronomy, University of Waterloo, 200 University Avenue West, Waterloo, Ontario, Canada N2L 3G1}

\begin{abstract}
Individual dark matter halos in cosmological simulations vary widely in their detailed structural properties such as shape, rotation, substructure and degree of internal relaxation.
Recent non-parametric (principal component) analyses suggest that a few principal components explain a large fraction of the scatter in halo properties. 
The main principal component is closely linked with concentration, which in turn is known to be related to the mass accretion history of the halo. 
Here we examine more generally the connection between mass accretion history and structural parameters. The space of mass accretion histories has principal components of its own. We find that the strongest two can be interpreted as the overall age of the halo and the acceleration or deceleration of growth at late times.
These two components only account for $\sim70$\%\ of the scatter in mass accretions histories however, due to the stochastic effect of major mergers. Relating structural parameters to formation history, we find that concentration correlates strongly with the early history of the halo, while relaxation correlates with the late history. We examine the inferences about formation history that can be drawn by splitting haloes into subsamples, based on observable properties such as concentration and shape at some final time. This approach suggests interesting possibilities, such as the possibility of defining young and old samples of
galaxy clusters in a rigorous, quantitative way, or testing the dynamical assumptions of galaxy formation models empirically. 
\end{abstract}


\maketitle

\section{Introduction}
\label{sec:introduction}
 Dark matter halos provide the framework for visible structure in the universe over a span of eight decades in mass, from the scale of rich galaxy clusters down to the scale of individual dwarf galaxies.  Mass is normally assumed to be the main determinant of a halo's baryonic contents, and analytic models such as the halo occupation distribution \citep[HOD --][]{Peacock, Seljak, Ma, White, Berlind, Cooray2002-halomodel}
 make this assumption explicitly. Yet the halos that form in cosmological simulations vary greatly in shape, concentration, spin, substructure and other structural properties. As simulations of increasing size and resolution provide a more and more detailed picture of halo properties, and as observational techniques including weak and/or strong gravitational lensing, X-ray and Sunyaev--Zel'dovich measurements \citep[e.g.][]{Corless2009-WeaklensingShape,Oguri2010-WeaklensingShape25,Sereno2011-StronglensingShape,Morandi2011-xraylensing}
reach a precision where they can determine structural properties such as shape and concentration reliably, it is important to understand how the structural features of a halo are interrelated, and what they can tell us specifically about its formation and evolution. 

Recently two groups, \citet{Skibba2011-PCA} and \citet{JeesonDaniel2011-PCAcorrelations} (S11 and J11 hereafter), have taken the important step of performing non-parametric principal component analyses of halo properties. Principal component analysis (PCA) searches for simplifying trends in a complex data set by finding the axes in a multi-dimensional parameter space that account for the largest fraction of the scatter. In the simplest case, it can uncover linear correlations in the data (e.g. fundamental lines or planes) and reveal hidden patterns or simplifications. The results of S11  and J11 agree on some basic aspects of halo structural properties. Overall, the scatter in halo properties spans a fairly high-dimensional space, with 4 principal components required to explain about 70\% of the scatter. Nonetheless, a few strong components emerge. 

In terms of structural parameters, the first (strongest) principal component is best traced by concentration. The strong correlation with concentration suggests that this first principal component is linked to the overall age of the halo, and the strength of the correlation with $z_{0.5}$ confirms this. The origin of halo concentration has been studied extensively since it is a crucially important factor in many calculations, including strong lensing \citep[e.g.][]{Broadhurst2005-stronglensing,Broadhurst2008-stronglensing2} and dark matter annihilation \cite[e.g.][]{Taylor2003-darkmatterannihilation}. Several analytic models have been developed over the years to explain concentration in terms of formation history \citep[e.g.][]{Bullock2001-concentrationMAH, Wechsler2002-MAH, Zhao09}, so the link between formation history and this particular structural property is fairly well understood. Older systems, that is systems that had already assembled most of their mass into one or a few progenitors at early times, are more concentrated on average, although the exact details of the connection between age and concentration vary from one analytic model to another.
 
If the first principal component of halo structural properties is thus linked to age, we might naturally ask what the others correspond to. Are shape, spin or relaxation also related to the formation history, and if so how? To put this question in a quantitative framework, we first have to decide how to describe the ``formation history" itself -- what should we take this to mean exactly, given the complex set of merger and accretion events through which halos form?

We can take the mass accretion history (MAH) as a starting point, defining this as the function ${\mathcal{M}}(z) \equiv M(z)/M(0)$ which describes the mass of the main progenitor of a halo as a function of redshift, normalized to the value at $z = 0$ \citep{vandenBosch2002-MAH}. Since ${\mathcal{M}}(z)$ is a continuous function of a real variable, it contains an arbitrarily large amount of information about the history of a halo; equivalently, describing the MAH fully means 
specifying values of ${\mathcal{M}}(z)$ at an infinite set of redshifts. To characterize the MAH in simpler terms, we can turn once again to principal component analysis, approximating each MAH as a vector of values ${\mathcal{M}}(z_i)$ corresponding to the MAH evaluated at a finite, fixed set of redshifts $z_i$. PCA of these vectors can then tell us whether mass accretion histories are well described by a set of basis functions characterzed by a single parameter, as suggested by \citet{vandenBosch2002-MAH} and \citet{Wechsler2002-MAH}, or whether they require two or more variables to explain their diversity, as suggested by \citet{Tasitsiomi} or \citet{McBride2009-MAH}. 

Assuming a few principal components capture the main features of a halo's MAH, this will allow us to study correlations between structure and history in a well-defined and quantitative way. We note that this is only a first step towards understanding structure in terms of growth history; the MAH does not contain all the information about a halo's past by any means, and an alternate approach is to study the origin of a particular property such as shape or relaxation in detail, considering the physical processes involved \citep[e.g.][]{VeraCiro2011-ShapeAquarius,Power2011-CorrelationsMAH}.

In this paper, we generate halos in a set of cold dark matter (CDM) simulations covering three different mass scales. We record the MAH of each well-resolved halo and decompose the space of MAHs, taken as vectors of values ${\mathcal{M}}(z_i)$, into principal components. These components provide a non-parametric description of the MAH and clarify its basic properties. We also record the structural properties of each halo at the present-day. We study the correlations between these properties themselves (reproducing the trends found by S11 and J11), and between structural properties and the principal components of the merger history. Finally, we discuss an application of our results, showing how samples of halos selected by concentration or shape will have systematically different formation histories. In this way, the observable properties of groups or clusters of galaxies can be used to infer their (unobservable) formation history.
 
The outline of the paper is as follows. Section \ref{sec:measuring_halo_properties} describes our n-body simulations and group finding, and how the structural properties of halos are defined and measured. Section \ref{sec:MAHs} analyses the vector space of MAHs, decomposing them into principal components. In Section \ref{sec:structural_cor}, we then analyze the correlations between structural components, and derive principal components in this second vector space. 
In Section \ref{sec:application} we discuss applications of this work. We summarize our results in Section \ref{sec:summary}.

Throughout the paper, we consider a WMAP7 cosmology with parameters $\Omega_M = 0.27$, $\Omega_\Lambda=0.73$, $\Omega_b=0$ (in our CDM-only simulations), $H_{0} = 72$ $\t{km} \t{s}^{-1} \t{Mpc}^{-1}$, $n_s = 0.96$, and $\sigma_8 = 0.80$.

\section{Measuring Halo Properties}
\label{sec:measuring_halo_properties}

\subsection{Numerical Simulations}
Our simulations were performed using the massively parallel n-body code \texttt{Gadget2} \citep{Springel2005-Gadget2}, with cosmological initial conditions generated by  initial condition code \texttt{Grafic2} \citep{Bertschinger2001-Grafic2}. Three different boxsizes were simulated with $N_p = 512^3$ particles each, producing halo catalogs that vary in mass by a factor of 8 from one simulation to the next.

The softening length used in the simulations was $\epsilon = \min(\epsilon_{\t{com}},\epsilon_{\t{max}}/a)$, where $\epsilon_{\t{com}}$ is a comoving softening length and 
$\epsilon_{\t{max}}$ is a  (maximum) physical softening length.
The softening lengths and other simulation parameters are listed in Table \ref{tab:nbody_simulation_parameters}.
\begin{table}
\caption{Simulation Parameters}
\centering
\begin{tabular}{ c c c c c c }
\hline
Name & $m_p$ & $L_{\text{box}}$ & $z_{\text{start}}$ & $\epsilon_{\text{com}}$ & $\epsilon_{\text{max}}$ \\
{} & $10^{10} M_{\odot} h^{-1}$ & Mpc $h^{-1}$ & & kpc $h^{-1}$ & kpc $h^{-1}$ \\
\hline\hline
Sim60 & $0.01206$ & 60 & 81.1 & 50 & 5 \\
Sim120 & $0.09648$ & 120 & 67.3 & 110 & 11 \\
Sim240 & $0.77184$ & 240 & 55.2 & 240 & 23 \\
\hline
\end{tabular}
\label{tab:nbody_simulation_parameters}
\end{table}

In each simulation, we output 95 snapshots spaced logarithmically in cosmic scale factor $a$ between $z=8$ and $z=0$. 
To detect halos in each snapshot we use the Friends-of-Friends (hereafter FOF) algorithm \citep{Davis1985-FOF} with a linking length of $b=0.2$ times the mean inter-particle separation $L_{\text{box}} \text{N}^{-1/3}$. We also use the SMOOTH algorithm \citep{Stadel1995-SMOOTH} to compute the mean local density around every particle. (These tools can be found at http://www-hpcc.astro.washington.edu/tools/tools.html.) 
From the initial FOF catalog we select all halos with 1000 particles or more, reducing the number of usable FOF groups 
to $1560$, $1580$, and $1532$ respectively for simulations Sim60, Sim120, and Sim240.
This gives us a total sample of $4672$ well-resolved halos.

In a sample defined solely by FOF linking, many groups will be multi-component systems caught in the act of merging.
This can introduce significant scatter in the measured structural properties of halos. We can flag unrelaxed systems
by using the minimum value of $\chi_{\rm NFW}^2$ (see Eqn.~ \eqref{eqn:chi_square_nfw} below) as a goodness-of-fit indicator.
Alternately, the asymmetry of the mass distribution can also be used to distinguish relaxed and unrelaxed systems.
The center of each halo is taken to be the position of the particle with the highest local mean density within the 
FOF group, $\vec{x}_{\rho}$, as centering on this point generally appears to produce smoother fits to the density profile 
than if we use the center of mass of the group, $\vec{x}_{\t{CoM}}$. 
The offset between $\vec{x}_{\rho}$ and $\vec{x}_{\t{CoM}}$ relative to the size of the halo is then an measure of asymmetry or relaxedness.
Defining $x_{\text{off}} \equiv |\vec{x}_{\rho} - \vec{x}_{\t{CoM}}|/r_{\rm vir}$, we select from our full sample a sub-sample of $3290$ relaxed halos 
for which $x_{\text{off}} < 0.07$ and $\chi_{\rm NFW}^2 < 0.5$. These correspond to the criteria used by \cite{Maccio2007-ConcentrationSpinShape} and S11 
(our $\chi^2_{\rm NFW}$ is their $\rho_{\rm rms}$). We note that they are fairly inclusive, defining 70\%\ of our sample as relaxed; 
\cite{Power2011-CorrelationsMAH} suggest using the stricter criterion
$x_{\text{off}} < 0.04$, for instance, to select relaxed systems.
In what follows, we will consider results both for the full sample and for the relaxed sub-sample.

\subsection{Density Profiles}

To determine the spherically-averaged density profile, we bin the halo particles radially in equal-sized bins, starting from the center of the halo (i.e.~$\vec{x}_{\rho}$) and proceeding outwards until we reach the virial radius $r_{\t{\rm vir}}$, defined by 
\begin{equation}
M_{\t{vir}} = (4\pi r_{\t{vir}}^3 /3)\Delta_{\rm vir,c}\rho_{\t{c}}\,,
\label{eqn:virial_mass_defn}
\end{equation}
where $M_{\t{vir}}$ is the virial mass of the halo, taken to be the FOF mass, and $\Delta_{\t{vir,c}} = 200$. We note that a fraction of the outer particles linked by FOF may lie outside $r_{\t{vir}}$, particularly at low redshift.

We fit a Navarro-Frenk-White \citep[][NFW hereafter]{Navarro1996-NFW} density profile to each halo over the radial range $\left[0.01 r_{\t{vir}},r_{\t{vir}}\right]$. We do not include the innermost 1\% of the virial volume in the fit, since resolution effects and softening may systematically alter the density profile there.
The NFW profile is
\begin{equation}
\rho_{\t{NFW}}(r) = \frac{\rho_s}{(r/r_s)(1+r/r_s)^2}\,,
\label{eqn:nfw_profile}
\end{equation}
where $r_\t{s}$ the scale radius  (the radius at which the logarithmic density slope is $-2$) and $\rho_\t{s}$ is the characteristic density. Best fit values of these parameters are obtained by minimizing the quantity
\begin{equation}
\chi_{\t{NFW}}^2 = \frac{1}{\t{N}_{\t{bins}}} \sum^{\t{N}_{\t{bins}}}_{i=1} \left[ \ln \rho(r_i) - \ln \rho_{\t{NFW}}(r_i) \right]^2\,,
\label{eqn:chi_square_nfw}
\end{equation}
where $\t{N}_{\t{bins}}$ is the number of bins used in the fit. 
This minimization technique is logarithmic rather than linear in order to give similar weight to the fitting near the centre and at the outer edges of the halo
\citep{Jing2000-concentration,Maccio2007-ConcentrationSpinShape}. 
The minimization code uses the Levenberg-Marquardt method to find the best fit values of $r_\t{s}$ and $\rho_\t{s}$.
As mentioned previously, we find that using the particle with the highest local density as the center of the halo produces good NFW fits to the density profile, while fits using the centre of mass have larger residuals. Bin size also affects our results somewhat;  after some experimentation we chose a bin size of 15 particles to measure the profile and determine halo concentration. 

\subsection{Concentrations}

Given an NFW density profile fit, the concentration parameter is defined as $c \equiv r_{\t{vir}} / r_{\t{s}}$. Here $r_{\t{vir}}$ is the virial radius, that is 
the radius around the halo center that encloses mass $M_{\t{vir}}$, as defined in Eqn.~\eqref{eqn:virial_mass_defn}.
Physically, concentration is a measure of the size or density of the central core relative to the size or density of the whole halo.

We could determine the concentration of a halo by measuring $r_{\t{vir}}$ and  $r_{\t{s}}$ separately and taking the ratio of the two. 
In practice, a slightly different technique appears to give more robust results.
Integrating the NFW profile with respect to radius out to $r_{\rm vir}$, the enclosed mass is given by
\begin{equation}
M_{\t{vir}} = 4 \pi \rho_s r_s^3 \left[ \ln(1+c) - \frac{c}{1+c} \right] 
\label{eqn:NFW_M_integral}
\end{equation}
Comparing the definition of the virial mass (Eqn.~\eqref{eqn:virial_mass_defn}) with 
Eqn.~\eqref{eqn:NFW_M_integral}, 
we can derive the non-linear relation
\begin{equation}
f(c) \equiv {1\over{c^3}}\left[ \ln(1+c) - \frac{c}{1+c} \right] = \left( {\Delta_{\rm{vir,c}}\rho_c}\over{3 \rho_s} \right)\,.
\label{eqn:NFW_c_eqn}
\end{equation}
Solving this equation then gives us the concentration $c$, as in \citep{Klypin2002-concentration,Gavazzi2003-concentration}.
We will use this second method throughout the rest of the paper, although we find the two methods for computing concentration are 
generally consistent with one-another to within 2--3\%. Typical errors in concentration are large (20--30\%) for our smallest halos, 
but drop below 10\%\ for halos with more than 3,000 particles.

\subsection{Mass Accretion Histories and Ages}
\label{subsec:MAH_age}
To determine the MAH for a given halo, we construct its merger tree, working backwards from the final snapshot at $z=0$.
The merger tree links the final halo to its progenitors at each previous redshift step.
To determine whether a halo $\mathcal{H}_{\t{later}}$ found at a later time is related to an earlier halo $\mathcal{H}_{\t{earlier}}$, we require that more than half of the particles in $\mathcal{H}_{\t{earlier}}$ be contained in $\mathcal{H}_{\t{later}}$.
If this condition is satisfied then we say that halo $\mathcal{H}_{\t{earlier}}$ is a parent halo (or progenitor) of the child halo $\mathcal{H}_{\t{later}}$. 
This method of defining parent halos restricts all earlier halos to have at most 1 child halo in the consecutive snapshot. 
The method does not restrict the number of parent halos a child halo may have, so in order to construct the MAH, we define the main parent to be the one with the largest contribution of particles to the child halo. This then produces a single sequence of parent halos in the merger tree, so we can assign a well-defined MAH to the halo in the final snapshot.

We fit our final MAHs with the 2-parameter model proposed by \cite{Tasitsiomi} and \cite{McBride2009-MAH}
\begin{equation}
\mathcal{M}(z) \equiv M(z)/M(0) = (1+z)^{\beta} e^{-\gamma z}\,,
\label{eqn:mcbride_mah}
\end{equation}
where $\beta$ and $\gamma$ are free parameters to be fit, and $M(0)$ is the mass of the halo at $z=0$. This form is a generalization of the earlier 1-parameter fit proposed by \citet{Wechsler2002-MAH}. The physical interpretation of the parameters $\beta$ and $\gamma$ can be seen by noting that
\begin{eqnarray}
-\frac{\t{d}}{\t{d}z} \ln \mathcal{M}(z) & = & -(1+z)^{\beta} e^{-\gamma z}  \left[ \beta (1+z)^{-1} - \gamma \right] \\
 & = & \gamma - \beta\ \ \ \t{for}\ z \sim 0 
\label{eqn:mcbride_mah_dz}
\end{eqnarray}
Thus $(\gamma - \beta)$ is the logarithmic growth rate of the halo in the limit $z \rightarrow 0$ (approaching from above).

We perform the model fits to each MAH by minimizing the quantity
\begin{equation}
\chi_{\mathcal{M}}^2 = \frac{1}{\t{N}_{\t{snaps}}} \sum^{\t{N}_{\t{snaps}}}_{i=1} \left[ {{M(z_i)}\over{M(0)}} - \mathcal{M}(z_i) \right]^2
\end{equation}
where $\t{N}_{\t{snaps}}$ is the number of snapshots for which the halo has existed in our sample.

\subsection{Shape and Triaxiality}
\label{subsec:shape_and_triaxiality}

In order to describe the shape of a halo, one can diagonalize its moment of inertia tensor 
\begin{equation}
\mathcal{I}_{ij}^{1} = \sum^{N}_{n=1} m_n \left( ||\vec{r}_n||^2 \delta_{ij} - r_{i,n} r_{j,n} \right)
\label{eqn:1st_MoI}
\end{equation}
where $r_{i,n}$ is the $i^{\t{th}}$ Cartesian component (relative to the halo center) of the $n^{\t{th}}$ particle of the halo. 
This will produce eigenvalues $\lambda_{i}^{\mathcal{I}^{1}}$ that are related to the relative axis lengths $\{a,b,c\}$ of the a halo (assuming an ellipsoidal mass distribution) by
\begin{eqnarray}
\lambda_{1}^{\mathcal{I}^{1}} = \frac{1}{5} M_{\t{halo}} \left( b^2+c^2 \right)
\label{eqn:eigenvalues_lambda_1_1st_MoI}\\
\lambda_{2}^{\mathcal{I}^{1}} = \frac{1}{5} M_{\t{halo}} \left( a^2+c^2 \right)
\label{eqn:eigenvalues_lambda_2_1st_MoI}\\
\lambda_{3}^{\mathcal{I}^{1}} = \frac{1}{5} M_{\t{halo}} \left( a^2+b^2 \right)\,.
\label{eqn:eigenvalues_lambda_3_1st_MoI}
\end{eqnarray}
With simple linear combinations of Eqn.~\eqref{eqn:eigenvalues_lambda_1_1st_MoI} - \eqref{eqn:eigenvalues_lambda_3_1st_MoI}, one can then obtain the relative axis lengths.

A more direct -- but equivalent -- method of obtaining the relative axis lengths $\{ a,b,c \}$ is to diagonalize the tensor describing 
the second moment of the mass distribution
\begin{equation}
\mathcal{I}^{2}_{ij} = \sum^{N}_{n=1} m_n r_{i,n} r_{j,n}
\label{eqn:2nd_MoI}
\end{equation}
The 3 eigenvalues $\lambda_{i}^{\mathcal{I}^{2}}$ from the diagonalization of the matrix in Eqn.~\eqref{eqn:2nd_MoI} will be the squares of the relative axis ratios, so the relative lengths of the principal axes can thus be determined as $\{a,b,c\} = \{ |\lambda_{1}^{\mathcal{I}^{2}}|^{\frac{1}{2}},|\lambda_{2}^{\mathcal{I}^{2}}|^{\frac{1}{2}},|\lambda_{3}^{\mathcal{I}^{2}}|^{\frac{1}{2}} \}$.
By convention, throughout the rest of the paper we will we re-label the axis lengths as necessary so that $a > b > c$. We note there are several other possible ways to measure shape -- see \citep{Zemp} for a discussion of the relative advantages of different techniques.

As well as taking ratios of the individual axis lengths, we will also consider the triaxiality index $T \equiv (a^2 - b^2)/(a^2 - c^2)$.
Oblate spheroids have $T=0$, while purely prolate halos have $T=1$.
Halos tend to have a skewed distribution of $T$ values, so we also consider the elongation parameter $E \equiv (b^2 + c^2)/(2 a^2)$, whose distribution is more Gaussian.

\subsection{Summary of  measured halo properties}
\label{subsec:chosen_halo_properties}

In studying correlations between halo properties, we will consider a large set of properties so as to overlap with the two recent PCA studies (S11 and J11). Clearly, not all the parameters listed below are independent. For example, we will assess several different definitions of shape, but they all ultimately depend on the two axis ratios $c/a$ and $b/a$. Our main goal will be to reproduce the patterns seen in S11 and J11, and then relate them to aspects of the MAH. Any redundant parameters will appear as strong correlations in our analysis, and can be combined or ignored accordingly.

\begin{enumerate}
\item {\bf Virial Mass $M_{\text{vir}}$ } \\
This is the total mass of halo within a virial radius $r_{\text{vir}} = r_{200,c}$ of the halo centre, which is taken to be the particle of highest local density within the linked group found by FOF.
The virial radius is defined such that the halo has an average density of $200 \rho_c$.

\item {\bf Concentration $c_{\text{200}}$ } \\
The concentration $c \equiv r_{\rm vir}/r_{s}$ is determined using Eqn.~\eqref{eqn:NFW_c_eqn}, as explained above.

\item {\bf Formation Redshift $z_{x}$ for $x \in (0,1)$ } \\
This is the redshift at which the MAH reaches ${\mathcal{M}}(z) \equiv M(z)/M(0) = x$. The most common example is $z_{0.5}$, at which $M(z_{0.5}) = 0.5 M(0)$. 

\item {\bf Mass Fraction History $(M/M_0)_{z}$ for $z > 0$ } \\
The mass fraction history is a complementary measure of age, equal to 
${\mathcal{M}}(z)$ for a specified value of $z$.

\item {\bf Relaxedness $x_{\text{off}}$} \\
This is the distance between the center of mass of the halo and the particle of highest local density, 
divided by the virial radius of the halo $r_{\text{vir}}$.
It is expected that relaxed halo without any recent mergers will have a small value for $x_{\text{off}}$. 
Throughout this paper, we will use $x_{\text{off}} < 0.07$, $\chi^2_{\rm NFW} < 0.5$ as a relaxedness condition.

\item {\bf Triaxiality $T$ } \\
Defined as $T = (a^2 - b^2)/(a^2 - c^2)$ where $a>b>c$ are the lengths of the principal axes of the halo. $T$ measures the prolateness or oblateness of a halo. 
Spherical and sausage-shaped halos have $T \approx 1$, while 
disk-shaped halos have $T \approx 0$.

\item {\bf Elongation $E$ } \\
This is defined as $E = [(b/a)^2 + (c/a)^2]/2$. The distribution of values of $E$ is more Gaussian than the distribution of triaxiality values.

\item {\bf Sphericity $c/a$} \\
This is a measure of the sphericity of the halo, dividing the shortest axis length $c$ by the major axis length $a$.

\item {\bf Spin $\lambda$ } \\
To provide a measure of rotation, we use a dimensionless spin parameter $\lambda \equiv J_{\text{vir}}/( 2 G M^3_{\text{vir}} r_{\text{vir}})^{1/2}$ where $J_{\text{vir}}$, $M_{\text{vir}}$, and 
$r_{\text{vir}}$ are the total angular momentum, virial mass, and virial radius, respectively.
Discussion of a different spin parameter definition can be found in \cite{Maccio2007-ConcentrationSpinShape}.

\item {\bf Environment $D_{n,f}$ for $n \in \mathbb{Z}$ and $f > 0$ }\\
The dimensionless environment parameter $D_{n,f}$ is defined as distance of the $n^{\text{th}}$ nearest halo that has a virial mass greater than $f \dot M_{\text{vir}}$, divided by its virial radius $r_{\text{vir}}$.
This definition is physically motivated because it scales as the tidal force to the $-1/3$ power \citep{Haas2011-Environment}.

\end{enumerate}

\section{Analysis of the Mass Accretion Histories}
\label{sec:MAHs}

\subsection{The ($\beta$,$\gamma$) Fit}

\begin{figure*}
\centering
\begin{tabular}{cc}
\includegraphics[height=0.25\textheight]{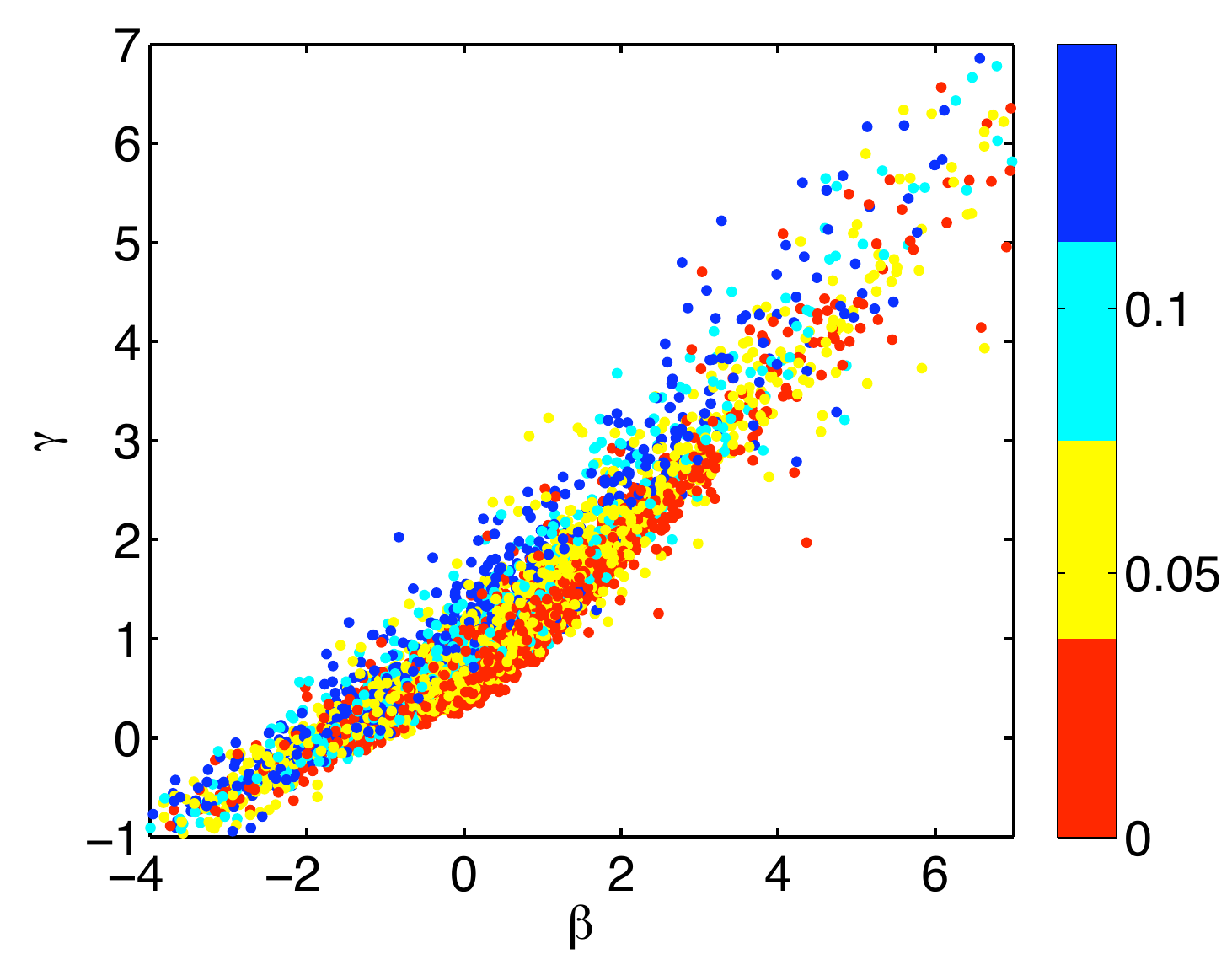} &
\includegraphics[height = 0.25\textheight]{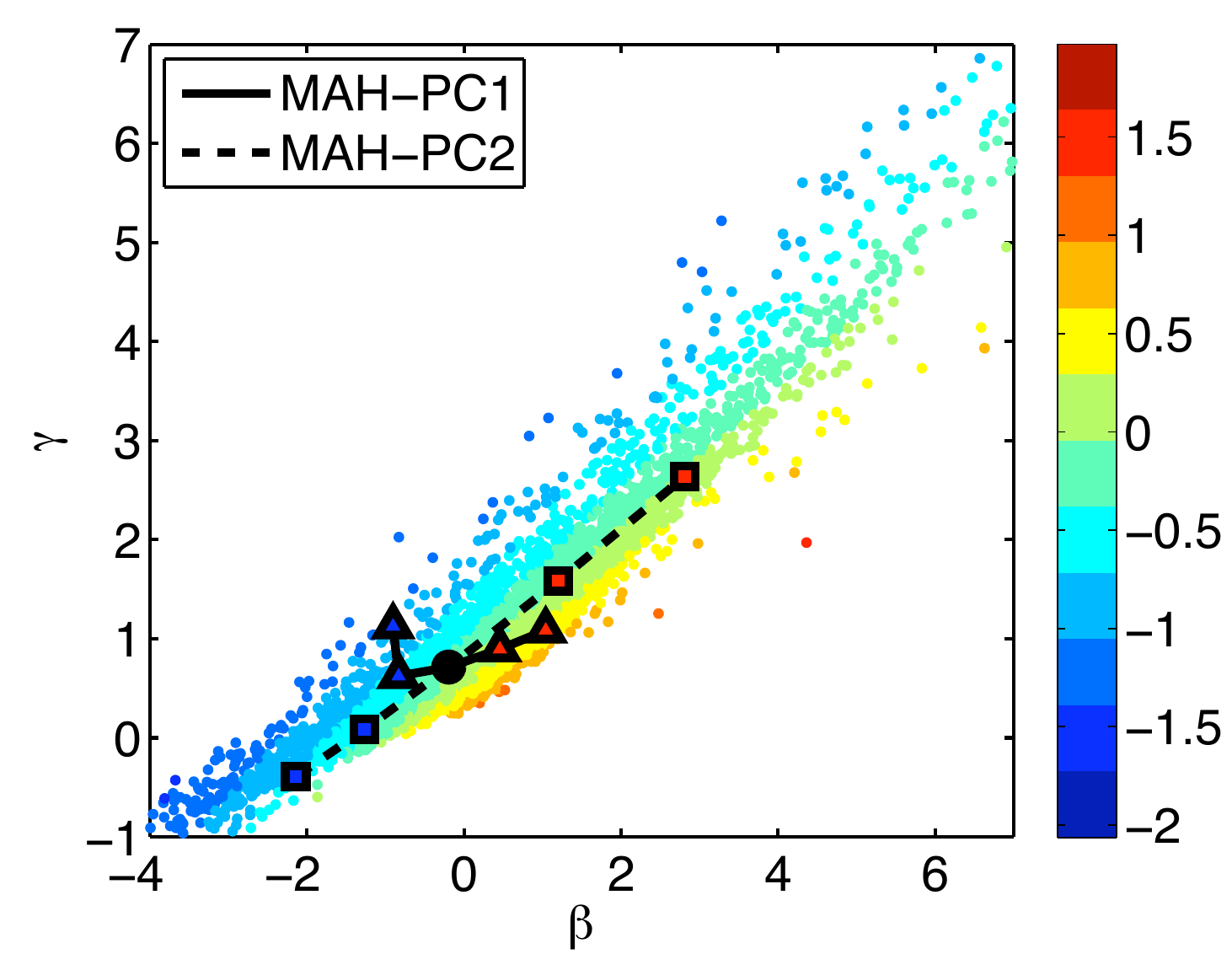}
\end{tabular}
\caption{Fitted values of $\gamma$ versus $\beta$ for 4672 halos with well-defined MAHs. The color scale indicates the value of relaxedness parameter 
$x_{\text{off}}$, with red points being more relaxed (left-hand panel), or the formation redshift $\text{log}_{10}(z_{0.5})$, with red points being older (right-hand panel). The large points on the right-hand panel indicate the values of $\beta$ and $\gamma$ used to fit the curves in 
Fig.~\ref{fig:mean_mah_pca_std_away_plot1} (triangles -- PC1; squares -- PC2). }
\label{fig:mah_beta_gamma}
\end{figure*}

For each of our well-resolved halos, we fit the functional form in Eqn.~\eqref{eqn:mcbride_mah} to the MAH to determine values of $\beta$ and $\gamma$. Fig.~\ref{fig:mah_beta_gamma} shows the resulting distribution of parameter values. The colour coding indicates the value of the relaxedness parameter
$x_{\text{off}}$ (left-hand panel), or the formation redshift as $\log_{10}(z_{0.5})$ (right-hand panel). 

While this functional form of Eqn.~\eqref{eqn:mcbride_mah} produces a reasonable fit to most of our MAHs, we can see immediately that $\beta$ and $\gamma$ are not necessarily a natural choice of variables, as there is a strong correlation between them for most MAHs. Furthermore, these parameters are degenerate in many cases, with different combinations of values providing almost identical fits to the MAH, as discussed in \citep{Taylor2011-HalosInsideOut}.
A more interesting variable is the quantity $(\gamma - \beta)$, which corresponds to the logarithmic growth rate as $z \rightarrow 0$. Given the simplicity of the ($\beta, \gamma$) fit, this quantity also fixes the overall age of the system, as can be seen clearly from the color scale in the right-hand panel of Fig.~\ref{fig:mah_beta_gamma}.
We also indicate on the right-hand panel of Fig.~\ref{fig:mah_beta_gamma} 
the values of $\beta$ and $\gamma$ 
used to fit the curves in Fig.~\ref{fig:mean_mah_pca_std_away_plot1} (see below).

\subsection{Principal Component Analysis of the MAHs}
\label{sec:MAHPCA}

Principal Component Analysis (PCA) is a non-parametric  technique for decomposing a set of $N_P$ possibly correlated variables into an equal or smaller number of independent variables. Because the final number of variables may be smaller than the initial number, PCA has the ability to reduce the dimensionality of a data set and uncover hidden patterns.

Technically, PCA involves determining the eigenvectors (principal axes $\bf{PC}_i$) of the covariance matrix and their respective eigenvalues (principal components $\lambda_i$). Using the convention that principal components are arranged by magnitude such that $\lambda_1 > \lambda_2 > ... > \lambda_{N_P}$, the principal components $\lambda_i$ are the relative fraction of the total variance in the direction of the principal axes $\bf{PC}_i$.
Thus PCA provides us with principal axes $\bf{PC}_1$ pointing in the direction of greatest variance in the data, $\bf{PC}_2$ pointing in the direction 
of second greatest variance orthogonal to $\bf{PC}_1$, etc.
Because the individual properties analyzed by PCA may have different dimensionality and/or magnitude, it is important to transform the initial data 
such that the mean of each parameter is 0 and the variance is 1, assuring that all the input fields will be treated equally.
Standardizing the variables in this way will also scale the principal components to sum to $N_P$, so the relative contribution
of the principal component $\lambda_i$ to the variance is the fraction $\lambda_i / N_P$.

We can use PCA to derive a more fundamental decomposition of the MAHs, taking as the input data vectors of values 
$\mathcal{M}(z_i)$ for a specified set of $z_i$. The resulting principal axes will then capture the basic shape variations of the MAHs. 
For this analysis we consider a subset of MAHs for $817$ halos extracted from Sim120 that are in the well-resolved sample 
(1000 particles or more, corresponding to masses of $10^{12} M_{\odot}$ or more) at $z = 0$, and have parents detected above 
the basic FOF limit of 200 particles or more (corresponding to masses of $2\times 10^{11} M_{\odot}$ or more) 
for at least 70 outputs (i.e.~going back to $z\sim 4$).   

\begin{figure*}[t]
\centering
\begin{tabular}{cc}
\includegraphics[height=0.25\textheight]{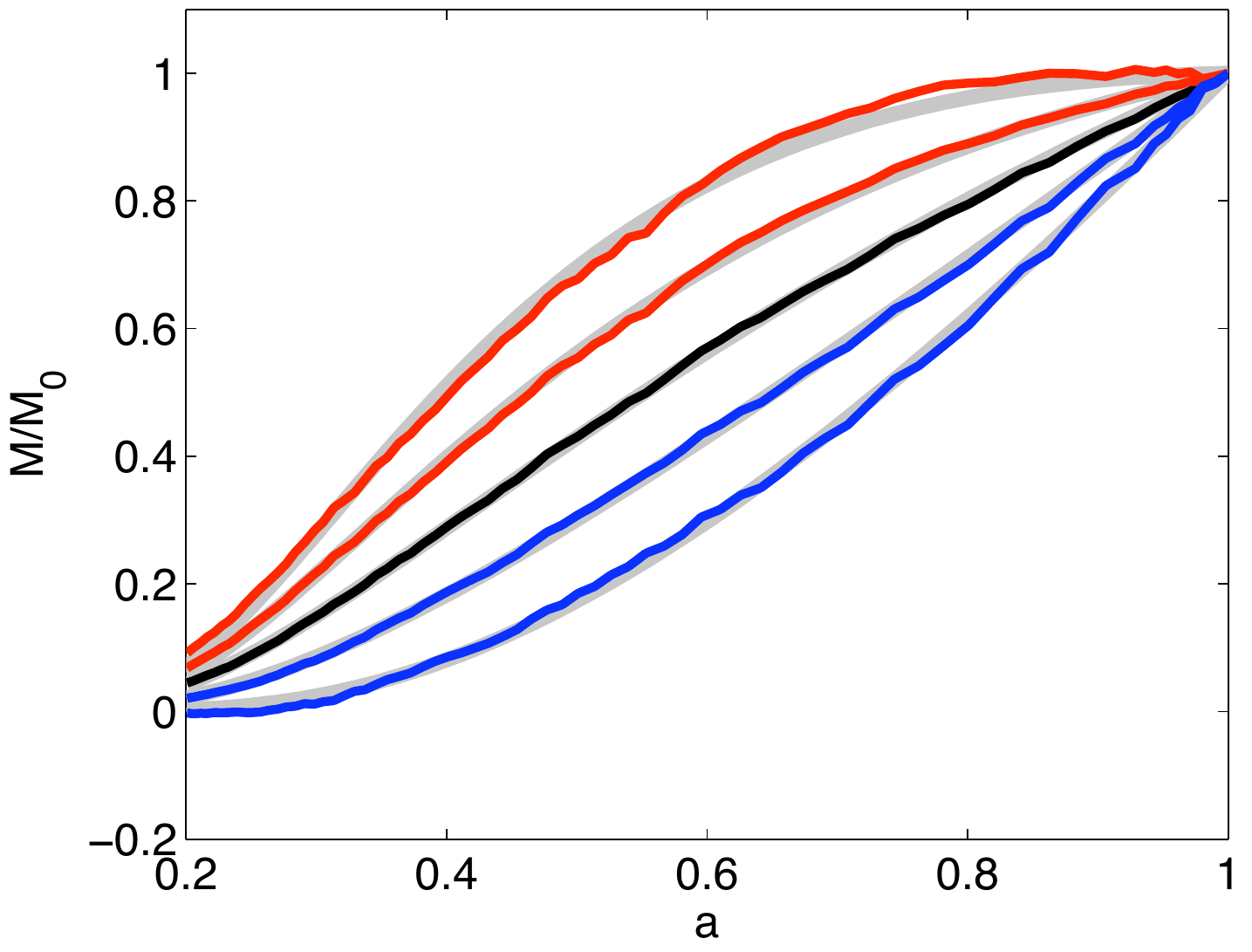} &
\includegraphics[height=0.25\textheight]{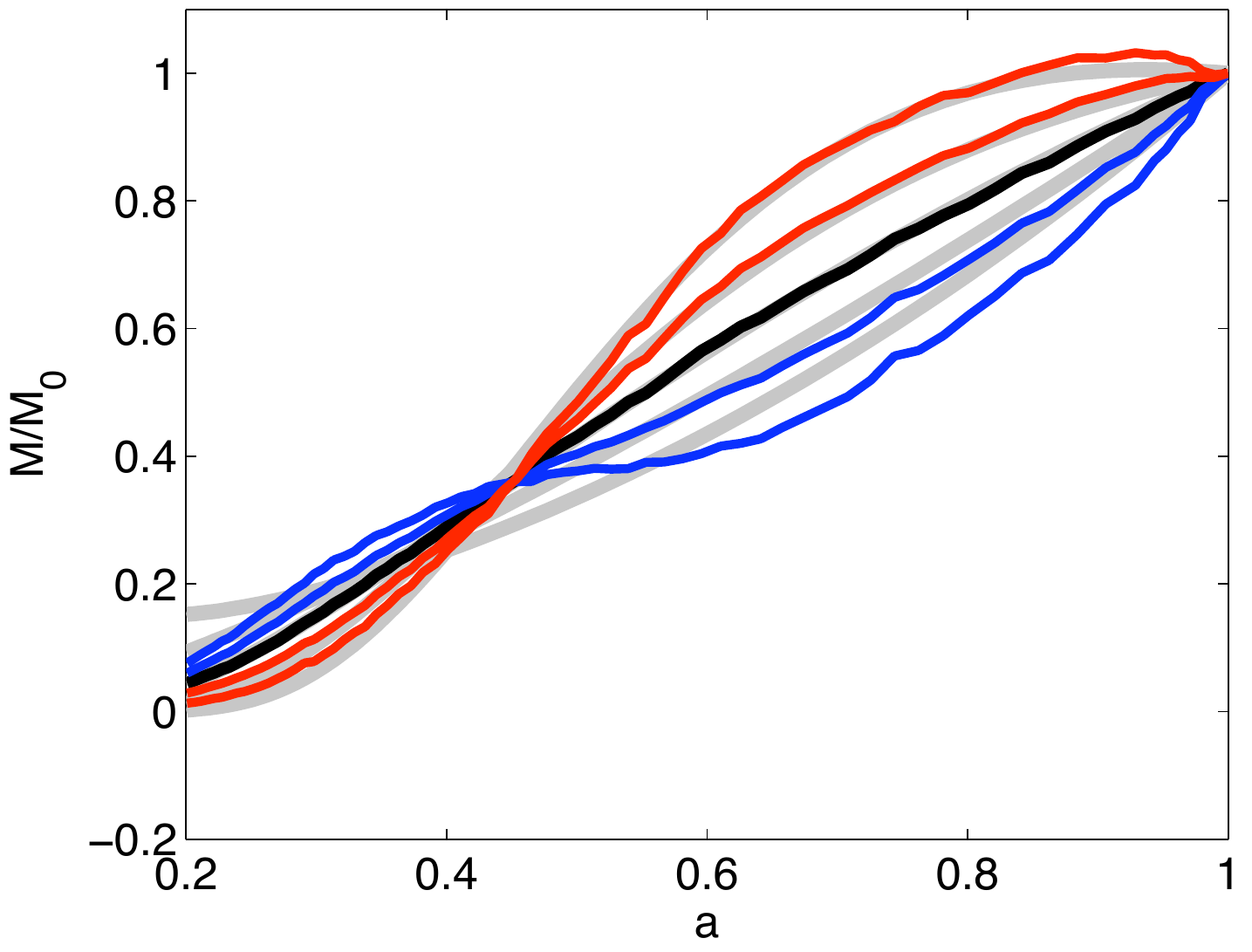}
\end{tabular}
\caption{The mean MAH for $817$ halos from Sim120 (black solid line), and its variation as one moves $\pm 1$ and $\pm 2$ standard deviations along the first (left-hand panel) and second (right-hand panel) principal component axes. In the left-hand panel, the upper (red) curves represent older halos, and the lower (blue) curves represent younger halos. In the right hand panel, they represent halos whose accretion rate has accelerated or decelerated close to $z = 0$, respectively. The gray lines indicate McBride fits to the curves.}
\label{fig:mean_mah_pca_std_away_plot1}
\end{figure*}

Fig.~\ref{fig:mean_mah_pca_std_away_plot1} shows the mean MAH for these systems (central black curve), plotted in terms of the scale factor $a$. The first principal component ({\bf MAH-PC}$_1$), which accounts for 51\% of the total variance in the MAHs, represents the most important variations away from this average. We can determine the nature of {\bf MAH-PC}$_1$ by adding or subtracting from the mean MAH the first principal component vector, normalized to 1 or 2 times the r.m.s. scatter along the {\bf MAH-PC}$_1$ axis. The resulting curves, illustrating $\pm$1--2 $\sigma$ deviations along the {\bf MAH-PC}$_1$ axis, are indicated by the upper (red) and lower (blue) curves in the left-hand panel of Fig.~\ref{fig:mean_mah_pca_std_away_plot1}. Clearly, {\bf MAH-PC}$_1$ corresponds to the overall age of the halo, with the upper (red) curves indicating older systems and the lower (blue) curves indicating younger systems. Both the mean and the $\pm$1--2 $\sigma$ deviations are all well fit by the McBride model (gray curves). The best fit values of $\beta$ and $\gamma$ are plotted as triangles on the right-hand panel of Fig.~\ref{fig:mah_beta_gamma}. Roughly speaking, they are spread out in the $(\gamma+\beta)=$ constant direction, but the precise positions appear complicated in this parameter space. This may be the result of the degeneracies in the ($\beta$, $\gamma$) fit discussed earlier.
  
The second principal component, {\bf MAH-PC}$_2$, accounts for a further 17\% of the variance. 
The right-hand panel of Fig.~\ref{fig:mean_mah_pca_std_away_plot1} shows the effect 
of $\pm$1--2 $\sigma$ deviations away from the mean along the {\bf MAH-PC}$_2$ axis. 
This variation is more complicated than that of {\bf MAH-PC}$_1$, but roughly speaking it represents acceleration or deceleration in the MAH. 
The ($\beta$, $\gamma$) model is no longer a particularly good fit to these curves, particularly the ones that accelerate rapidly at late times. 
The relationship between the best fit values of the $\beta$ and $\gamma$ is much simpler in this case, 
as indicated by the squares on the right-hand panel of Fig.~\ref{fig:mah_beta_gamma}, but given the poor fit to the curves it is not clear whether this is significant.

We have examined the other principal components of the MAH. Generally speaking, they appear to be Fourier-like decompositions of the curve 
with increasing numbers of oscillations about the mean MAH. Given that individually they account for relatively little of the scatter
(9\%\ for {\bf MAH-PC}$_3$, 6\%\ for {\bf MAH-PC}$_4$, $<$ 5\%\ for subsequent PCs), we will not discuss them any further. We note however that the total power 
in PCs 3 or higher is almost 30\%, showing that MAHs are not particularly well fit  by smooth curves, presumably because of the large jumps produced 
by major mergers. 

\begin{figure}[t]
\centering
\includegraphics[height=0.25\textheight]{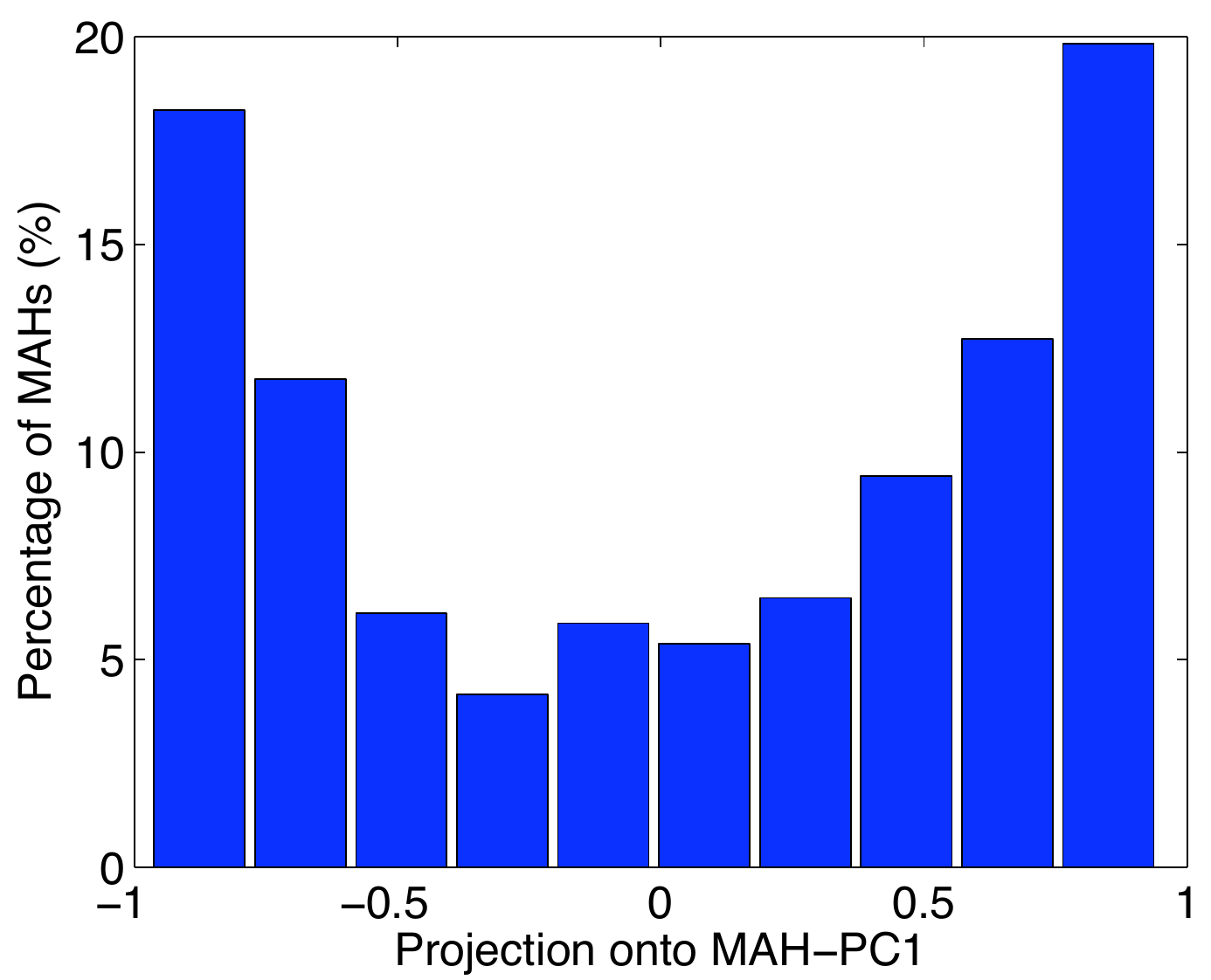}
\caption{The distribution of the {\bf MAH-PC}$_1$ coordinate value for $817$ well-resolved halos from Sim120.}
\label{fig:mean_mah_pca_std_away_plot2}
\end{figure}

It is also worth noting that the distribution of PCA vector components for individual MAHs is very far from a Gaussian scatter about the mean. 
Fig.~\ref{fig:mean_mah_pca_std_away_plot2} shows the distribution of the {\bf MAH-PC}$_1$ coordinate values for the 817 MAHs considered in 
Fig.~\ref{fig:mean_mah_pca_std_away_plot1}. 
Most MAHs have either high or low values of the {\bf MAH-PC}$_1$ component, suggesting that the mean MAH is not necessarily very 
indicative of individual MAHs, which tend to deviate strongly from the mean.

\subsection{Relation to Excursion Set Theory}

\begin{figure}[t]
\centering
\includegraphics[height=0.25\textheight]{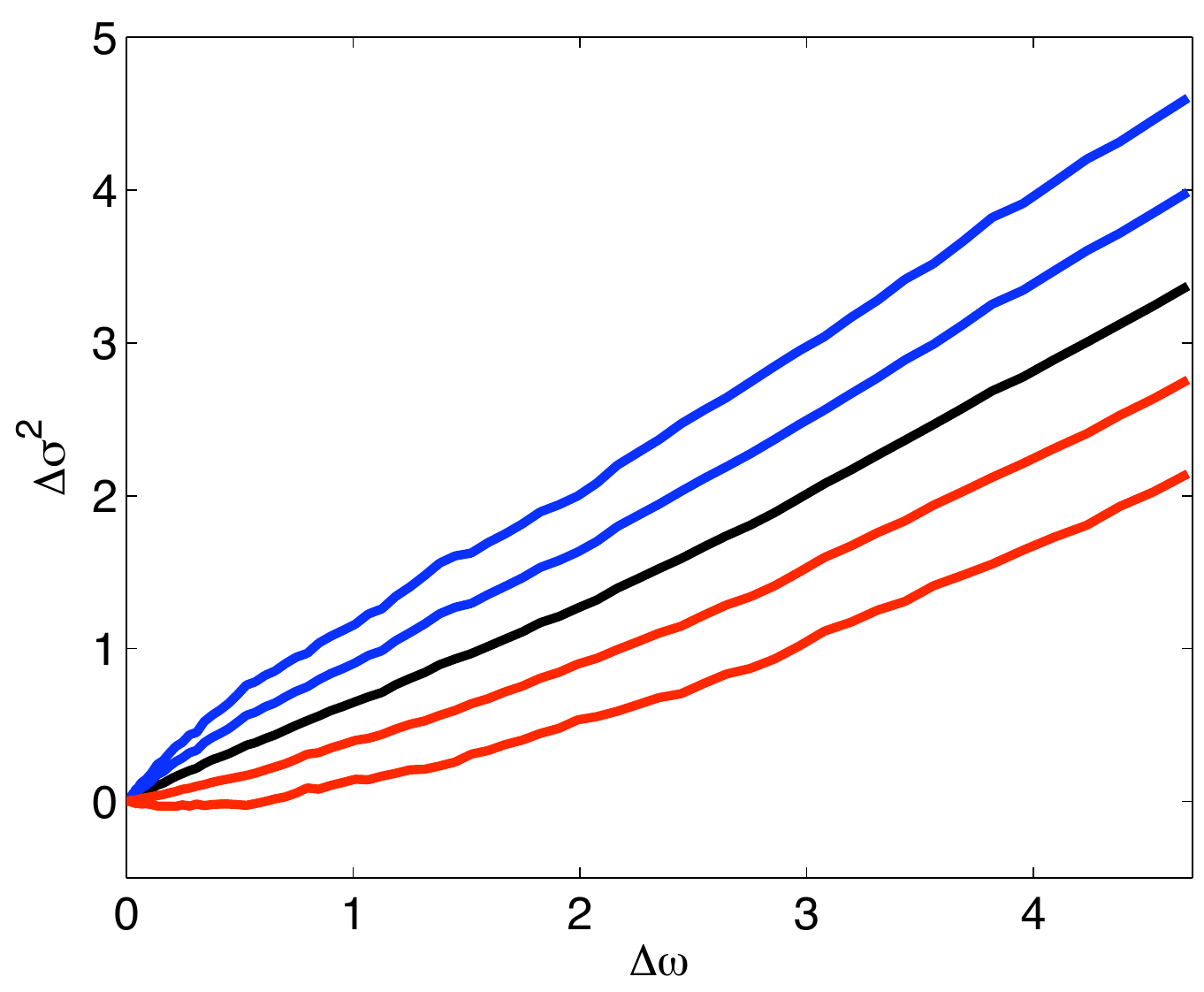}
\caption{The mean MAH (black line) and its variation by $\pm$1$\sigma$ and $\pm 2\sigma$ along the first  principal axis 
(upper/blue and lower/red solid lines), plotted in terms of the EPS 
variables $\Delta \omega$ and $\Delta\sigma^2$.}
\label{fig:mean_mah_pca_std_away_plot2_EPS}
\end{figure}

Finally, we can ask to what extent the basic shape of the mean MAH, or the variations of {\bf MAH-PC}$_1$ around it, reflect the statistics of halo growth 
predicted by excursion-set (or extended Press-Schechter -- EPS) theory. In excursion set theory, the growth of a halo can be considered a random 
walk in the space of variables $\sigma^2(M)$ and $\omega(z)$, where  $\sigma^2(M)$ is the variance of the primordial density field smoothed 
on a mass scale $M$, and $\omega(z) = \delta_c(z)$ is the height of the barrier for spherical collapse at redshift $z$ \citep{laceycole}. The distribution 
of jumps $\Delta\sigma^2$ over a redshift step corresponding to $\Delta\omega$ is a simply a Gaussian in $(\Delta\omega/\sqrt{\Delta\sigma^2})$ 
\citep[cf.][Eqn.~2.15]{laceycole}.
This suggests that MAHs should look much simpler in the coordinate system ($\omega$, $\Delta\sigma^2$). Fig.~
\ref{fig:mean_mah_pca_std_away_plot2_EPS} shows the mean MAH and the variations along the {\bf MAH-PC}$_1$ axis in ($\omega$, $\Delta\sigma^2$) coordinates. 
Sure enough, the different curves from Fig.~\ref{fig:mean_mah_pca_std_away_plot1} are now almost straight lines in these new coordinates. 
On closer examination, there is still a slight residual curvature to the lines. Tests with artificial MAHs suggest this is at least partly due to the difference 
between the variables $\Delta\sigma^2 = \sigma^2(M_j) - \sigma^2(M_0)$, which is plotted here, and $\Delta\sigma^2= \sigma^2(M_j) - \sigma^2(M_i)$, 
which is assumed in EPS theory, for a section of the trajectory going from $M_j$ to $M_i >  M_j$. It may also reflect from corrections introduced to EPS 
statistics by ellipsoidal collapse \citep{shethmotormen,shethtormen}. Overall, however, the transformation from ($\omega$, $\sigma^2$) 
to ($a$, $M$) coordinates accounts for most of the characteristic curvature of MAHs in ($a$, $M$) space. 
This is useful, as it gives an approximate way of predicting MAH distributions as a function of background cosmology.

\section{Analysis of Structural Properties}
\label{sec:structural_cor}

\subsection{Correlations Between Structural Parameters}

The simplest way to search for relationships between the structural properties of halos is to measure the correlation of one property with another. We will use the Spearman Rank coefficient as a quantitative measure of the correlation strength. The Spearman Rank coefficient, $r_\t{SRc} \in [-1,1]$, is a non-parametric measure of correlation which determines the strength of the relationship between two variables based on whether the data can be characterized by a monotonic function.
Perfect Spearman correlation corresponds to $r_\t{SRc} = 1$ (positive correlation) or $r_\t{SRc} = -1$ (negative correlation), while $r_\t{SRc} \approx 0$ indicates uncorrelated variables.0

The full correlation analysis of all 11 halo properties is shown in Appendix \ref{app:full_correlation_table}.  Broadly speaking, the strongest correlations are between age indicators such as the formation redshift $\log z_{0.5}$, and concentration $c$. Many parameters, such as $\log z_{0.5}$ and $\log[\mathcal{M}(0.8)]$, are nearly degenerate. Some weaker but physically significant correlations exist between mass and age indicators, and between shape and concentration. The former is easily understood as a consequence of hierarchical structure formation (more massive halos have formed more recently). The latter appears to be a newly discovered pattern.

Correlations between halo properties have already been analyzed by J11. 
In order to compare directly with their results, we also show, in  Appendix \ref{app:full_correlation_table}, a table similar to theirs. 
We omit substructure because we lack the resolution to detect it convincingly 
in many of our halos. Overall our results are almost identical to theirs;
concentration and age are the most fundamental parameters, with mass and shape following behind.
We find as they do that environment is not strongly correlated with any of the halo properties.
The only result that differs significantly is the correlation between the spin parameter 
$\lambda$ and concentration. It is not clear what the origin of this discrepancy is.

\subsection{Principal Component Analysis}
\label{subsec:principal_component_analysis}

We perform PCA, as described in Section \ref{sec:MAHPCA}, on the 11 halo properties described in Section \ref{subsec:chosen_halo_properties}. 
Our results are broadly consistent with the previous analyses of S11 and J11, but we include them here for completeness.
The strongest structural principal components ({\rm S-PCs}) with their respective principal axes are listed in Table \ref{table:PCA_results_all} for the full sample 
and Table \ref{table:PCA_results_relaxed} for the relaxed subsample. Numbers in bold indicate significant contributions to the principal component, taken here to be
coordinate values $|v_i| > 0.3$. We note that several of our parameters are almost degenerate, so our PC vectors have significant contributions from more parameters than in previous analyses (S11, J11). We summarize the properties of the four strongest PCs below. 

\begin{enumerate}
\item {\bf S-PC$_1$} includes concentration and the various age indicators ($z_{0.2}, z_{0.5}, (M/M_0)_{0.5}$) as its strongest contributors.
Mass and elongation also contribute significantly to PC1 for the relaxed sample; for the whole sample relaxedness $x_{\rm off}$ is more important than mass.
The overall weight of {\bf S-PC$_1$} and its composition are roughly in agreement with the PCA from S11 and J11; we verify that 
concentration, age, mass, and shape are seem to be related, fundamental parameters.
\item {\bf S-PC$_2$} contributes half to a third as much variance as {\bf S-PC$_1$}, and includes significant contributions from shape, age, and spin.
In fact, the contribution from shape parameters ($T$, $E$, $c/a$) is stronger than in {\bf S-PC$_1$}. 
This suggests that some aspect of the MAH of dark matter halos helps determine their shape at $z=0$.
\item {\bf S-PC$_3$} is then fairly distinct from the previous components, depending mainly on mass and spin. 
In the case of relaxed halos, environment also contributes at a lower but significant level to {\bf S-PC$_3$}.
\item {\bf S-PC$_4$} consists mainly of environment, showing that environment is a fairly independent parameter, 
or conversely that structural properties are largely independent of environment. For the relaxed subset there 
is some contribution from relaxedness and triaxiality, showing these do correlate with environment in relaxed systems.
\end{enumerate}

\input{pca_table2.tex}

\input{pca_table1.tex}

\subsection{The origin of halo shapes}
\label{subset:old}

\begin{center}
\begin{figure*}
\begin{center}
\subfigure[]{
\includegraphics[scale=0.55]{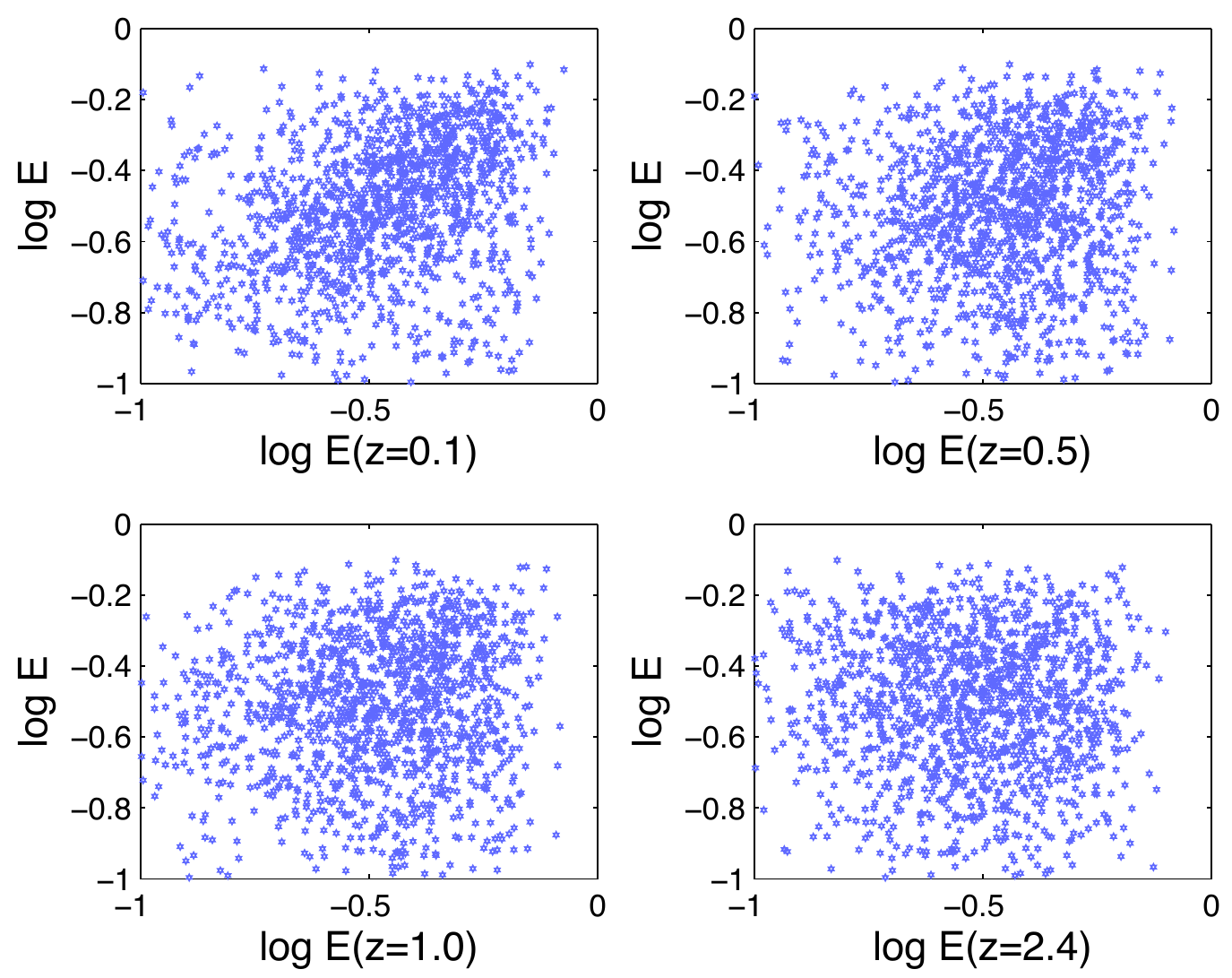}
}
\subfigure[]{
\includegraphics[scale=0.55]{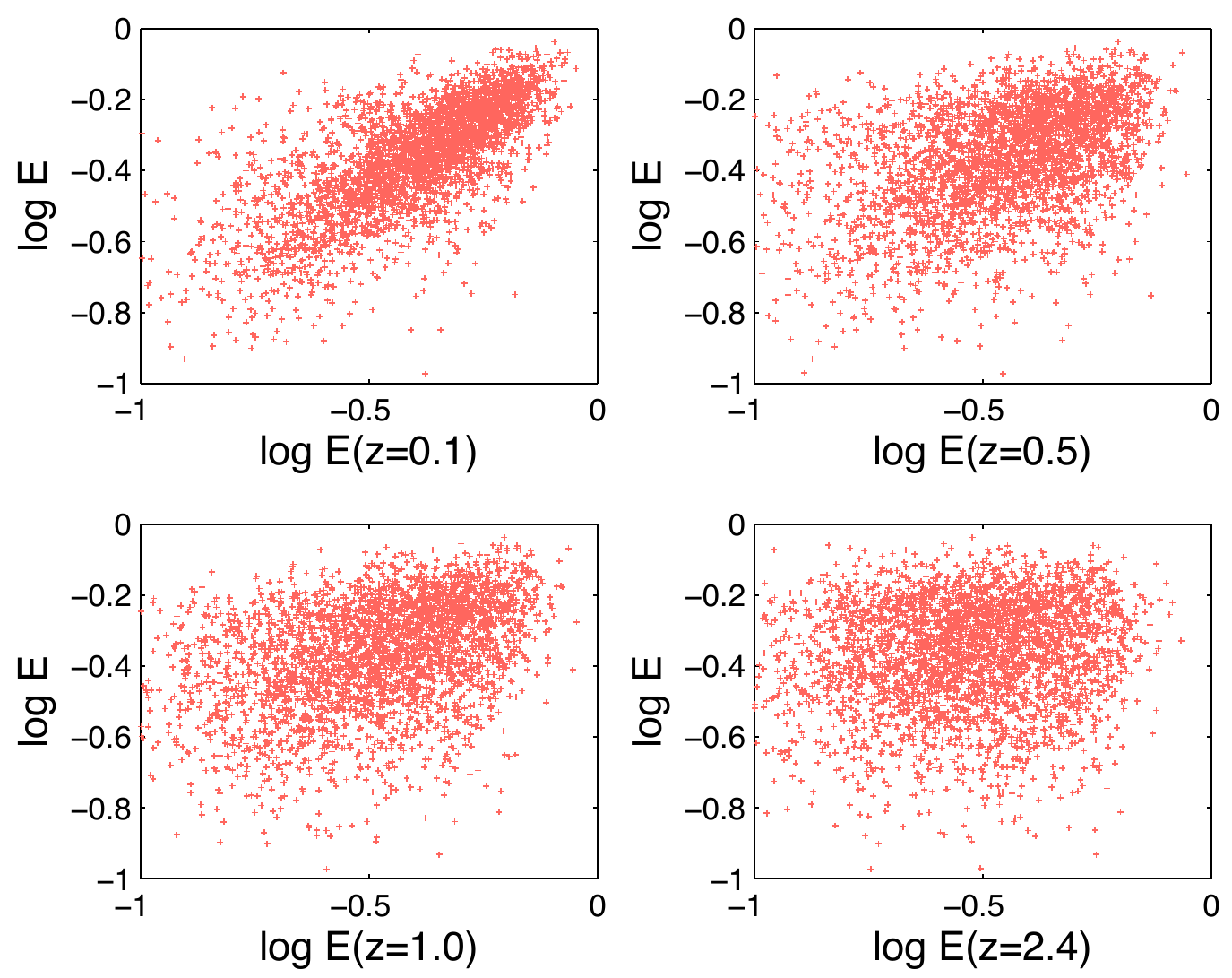}
}
\caption{Elongation $E$ of the halo versus elongation of the progenitor particle distribution at $z=0.1,0.5,1.0,2.4$ for relaxed (left) and unrelaxed (right) halos.}
\label{fig:fig5}
\end{center}
\end{figure*}
\end{center}

PCA on the structural properties of halos in Section \ref{subsec:principal_component_analysis}
indicated correlations between formation history and shape, particularly in (structural) PC2 
(cf.~Tables \ref{table:PCA_results_all} and \ref{table:PCA_results_relaxed}). There are various ways in which formation history could 
influence shape. One simple possibility is that the final shape of a halo is determined by the shape of the initial, uncollapsed matter from which it forms.
This connection would not be apparent in the MAH, since it does not record the shape of the merging material, but only the rate at which it merges. 
We can easily test the hypothesis that final shape tracks initial shape by measuring the latter in early timesteps, including all the particles which 
eventually merge to form a halo. 

Fig.~\ref{fig:fig5} shows how the elongation of a halo at $z=0$ correlates with the elongation of the particle distribution from which the halo formed, 
measured at a higher redshift, for the full (left-hand pane) and relaxed (right-hand panel) samples. There is a strong correlation between 
initial and final elongation looking back to recent redshifts, particularly for relaxed halos; at higher redshift or for less relaxed halos the correlation 
gradually weakens. Table.~\ref{table:oldshapes_correlation} gives the Spearman rank coefficients for the correlations, showing that they 
are significant back to $z=1$. The implication is that halo shape can trace the geometry of recent phases of the merger process. We note 
that \cite{VeraCiro2011-ShapeAquarius} have recently completed a much more thorough study of the nature and mechanisms by which 
shape evolves as a halo grows.

\begin{table}[h]
\caption{Correlation Between Final and Initial Elongation}
\centering
\begin{tabular}{| c | c | c |}
\hline
$z$ & $r_{E}$ {\text{(all halos)}} & $r_{E}$ {\text{(relaxed halos)}} \\
\hline
$0.1$  & 0.630 & 0.727 \\ 
\hline
$0.5$  & 0.338 & 0.444 \\ 
\hline
$1.0$  & 0.300 & 0.405 \\ 
\hline
$2.4$  & 0.082 & 0.128 \\ 
\hline
\end{tabular}
\label{table:oldshapes_correlation}
\end{table}

\subsection{Structural Properties Versus MAH-PCs}

Given our earlier non-parameteric decomposition of the MAHs into their own set of principal components in Section \ref{sec:MAHPCA}, we can examine how structural properties relate to the main features of the MAH. Fig.~6 shows how $z_{0.5}$, concentration and relaxedness $x_{\rm off}$ correlate with {\bf MAH-PC}$_1$ 
and {\bf MAH-PC}$_2$, for the full sample (left-hand panels) and the relaxed sub-sample (right-hand panels).  In the first row of panels, we see that the relationship 
between $z_{0.5}$ and the two MAH-PCs is complex. {\bf MAH-PC}$_2$ is fairly cleanly anti-correlated with $z_{0.5}$, in the sense that accelerating MAHs have lower values of $z_{0.5}$, but the relationship between {\bf MAH-PC}$_1$ and $z_{0.5}$ more complicated. For large values of $z_{0.5}$, the two parameters are reasonably
anti-correlated, but systems with low values of $z_{0.5}$ show a large scatter in $z_{0.5}$ at almost constant {\bf MAH-PC}$_1$ $\sim 0$. These are probably systems 
with recent major mergers in their MAH, although it is interesting that many are present even in the relaxed subsample (right-hand panels). This may indicate that the
criteria we have used to define the relaxed sample ($x_{\rm off} < 0.07$ and $\chi_{\rm NFW}^2 < 0.5$) are too inclusive. \cite{Power2011-CorrelationsMAH}, for instance, use the stricter criterion 
$x_{\rm off} < 0.04$.

Concentration (second row of panels) follows a similar pattern. At high concentration, concentration and {\bf MAH-PC}$_1$ are well correlated, while at low concentration
there is a large scatter at almost constant {\bf MAH-PC}$_1$. In this case, however, these are often unrelaxed systems; restricting ourselves to the relaxed sample, we see a cleaner correlation between concentration and {\bf MAH-PC}$_1$.  Overall, the distribution suggests a simple relation between concentration and {\bf MAH-PC}$_1$, which is occasionally disturbed by mergers. A similar analytic model relating concentration to the MAH has been proposed by \citet{Zhao09}; they distinguish phases of slow growth, where concentration gradually increases, from phases of rapid growth, when concentration is reset to a low value. Thus concentration should be a good age indicator for old, relaxed systems, but a poor one for low-concentration and/or unrelaxed systems.

Interestingly, the relaxedness parameter $x_{\rm off}$ (bottom row of panels) behaves quite differently. It is relatively uncorrelated with {\bf MAH-PC}$_1$, but is fairly well correlated with {\bf MAH-PC}$_2$, at least for the full sample (since the relaxed sample is cut on $x_{\rm off}$, this makes it hard to tell how much correlation is present in this case). For the full halo sample, unrelaxed systems have a systematically higher value of $x_{\rm off}$, and systems with larger values of $x_{\rm off}$ have experienced significantly more acceleration in their MAH.
\begin{center}
\begin{figure*}
\begin{center}
\subfigure{
 \includegraphics[width=0.22\textwidth]{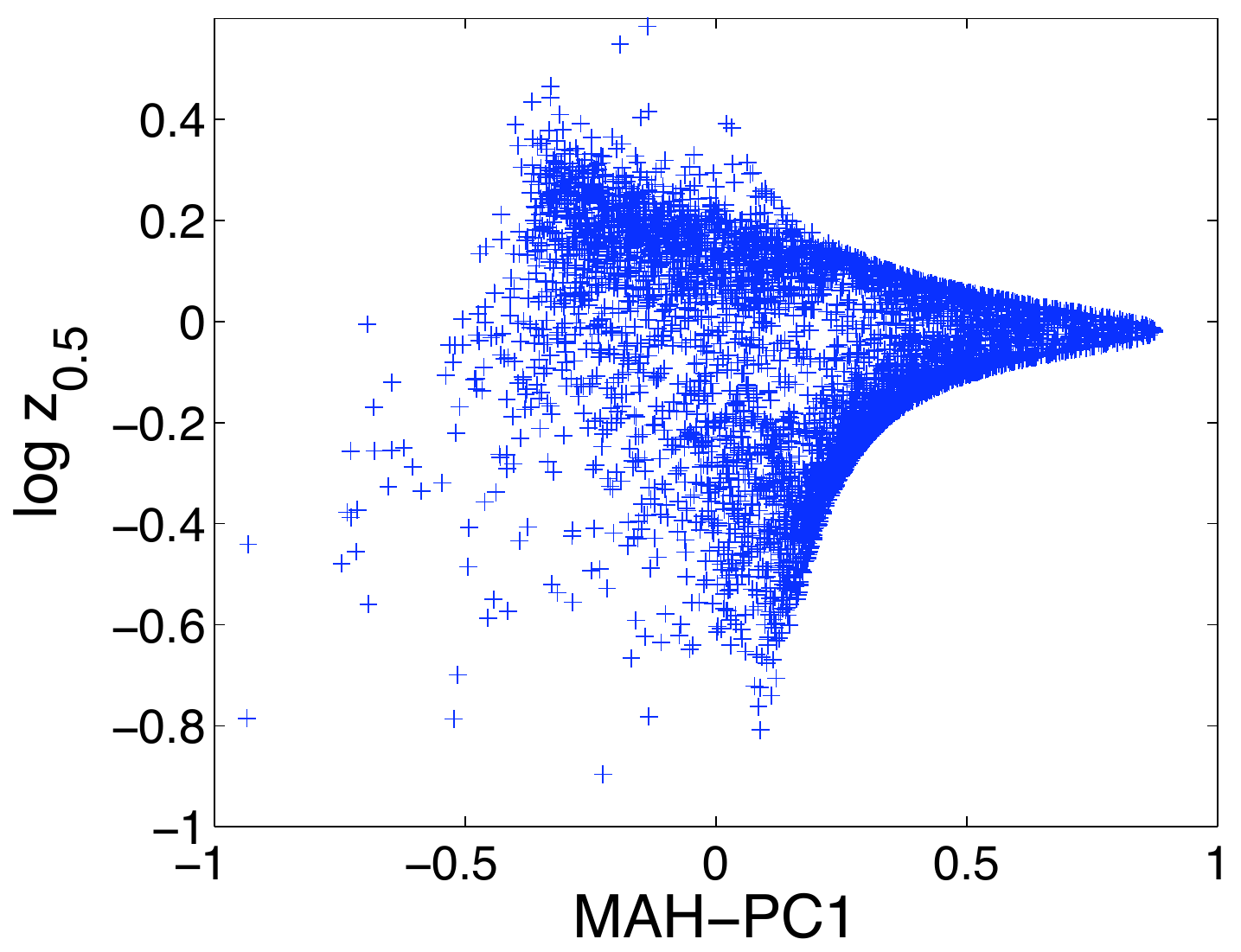}
 \label{subfig:mah_pcproj1_z05}
}
\subfigure{
 \includegraphics[width=0.22\textwidth]{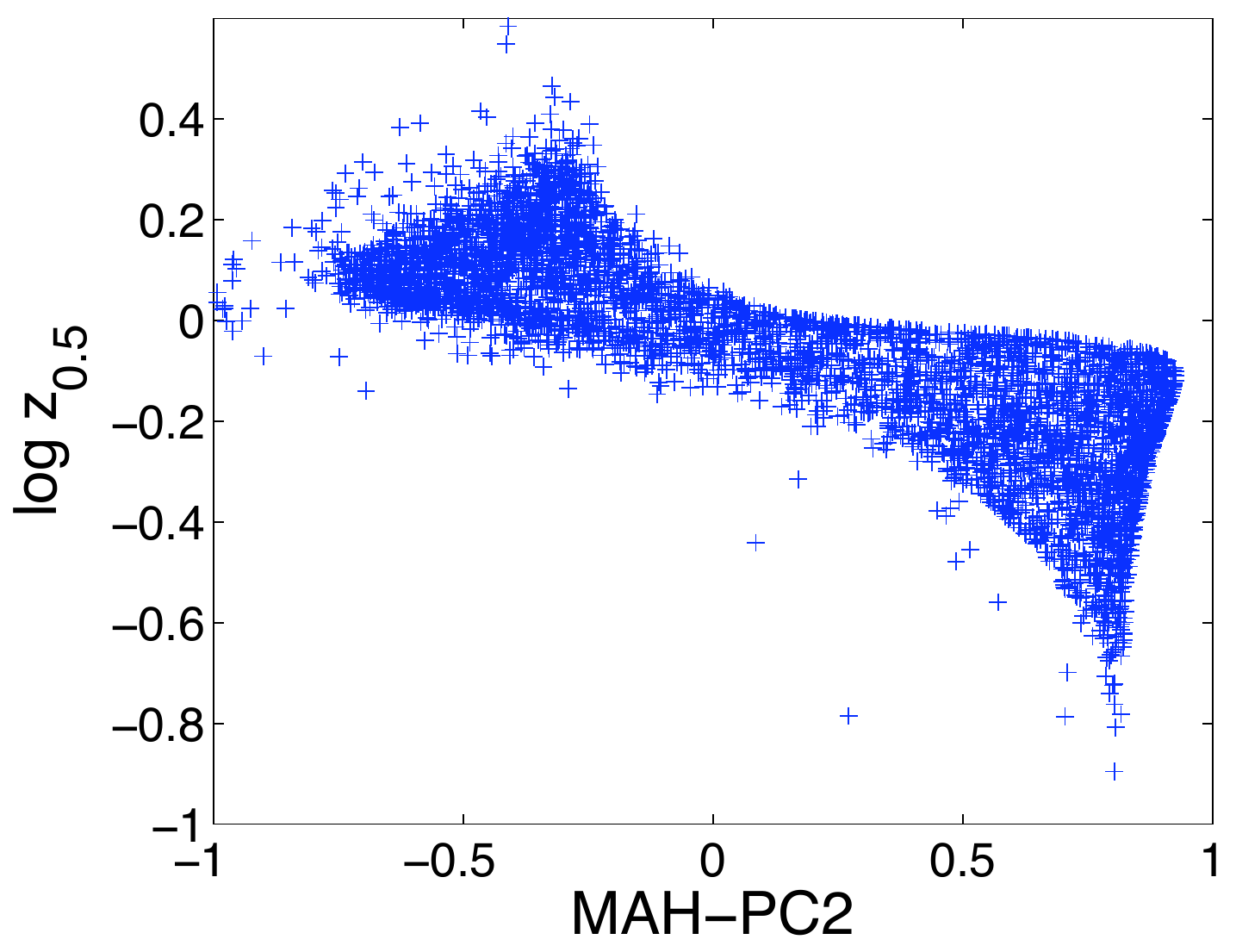}
 \label{subfig:mah_pcproj2_z05}
}
\subfigure{
 \includegraphics[width=0.22\textwidth]{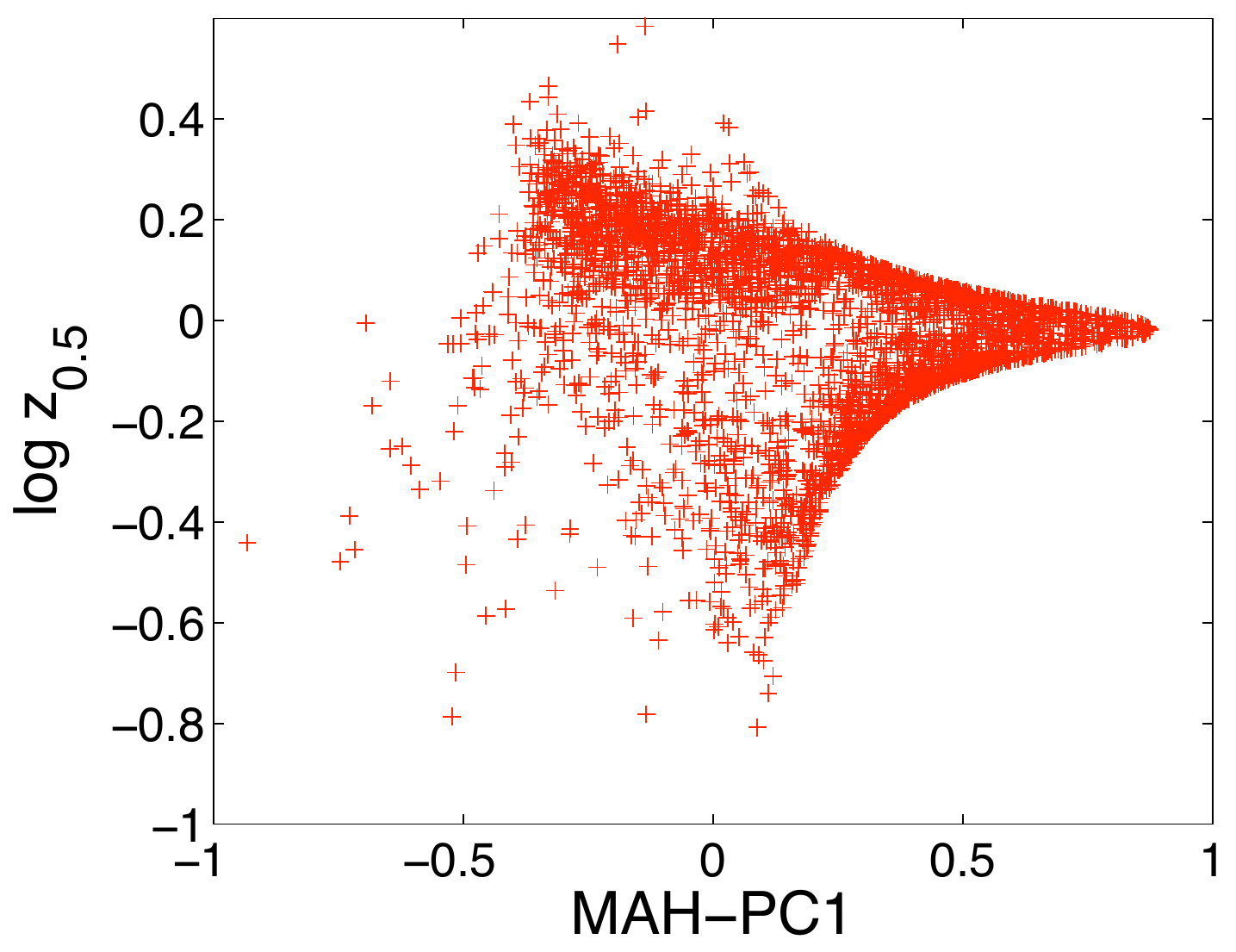}
 \label{subfig:mah_pcproj1_z05_relaxed}
}
\subfigure{
 \includegraphics[width=0.22\textwidth]{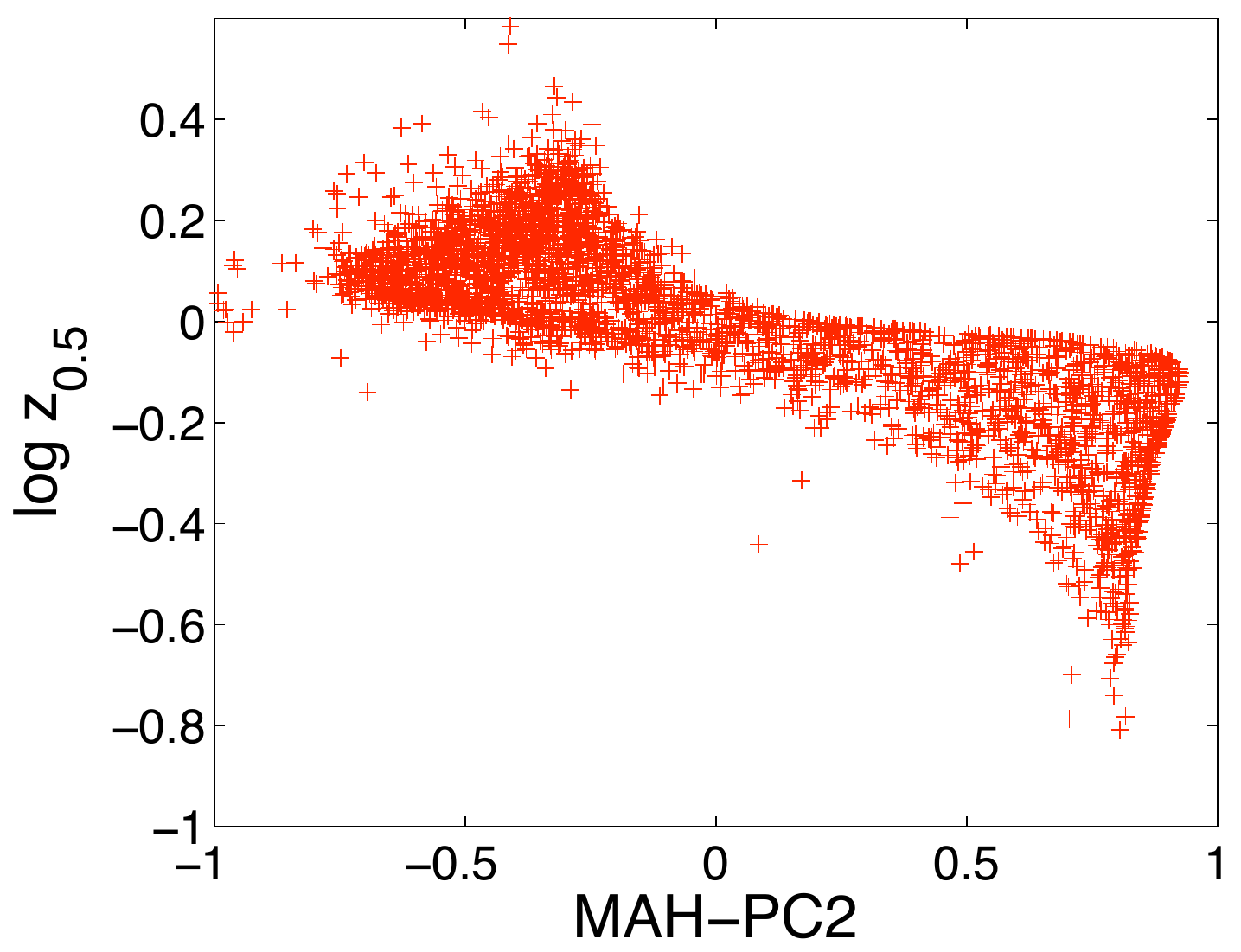}
 \label{subfig:mah_pcproj2_z05_relaxed}
 }
\subfigure{
 \includegraphics[width=0.22\textwidth]{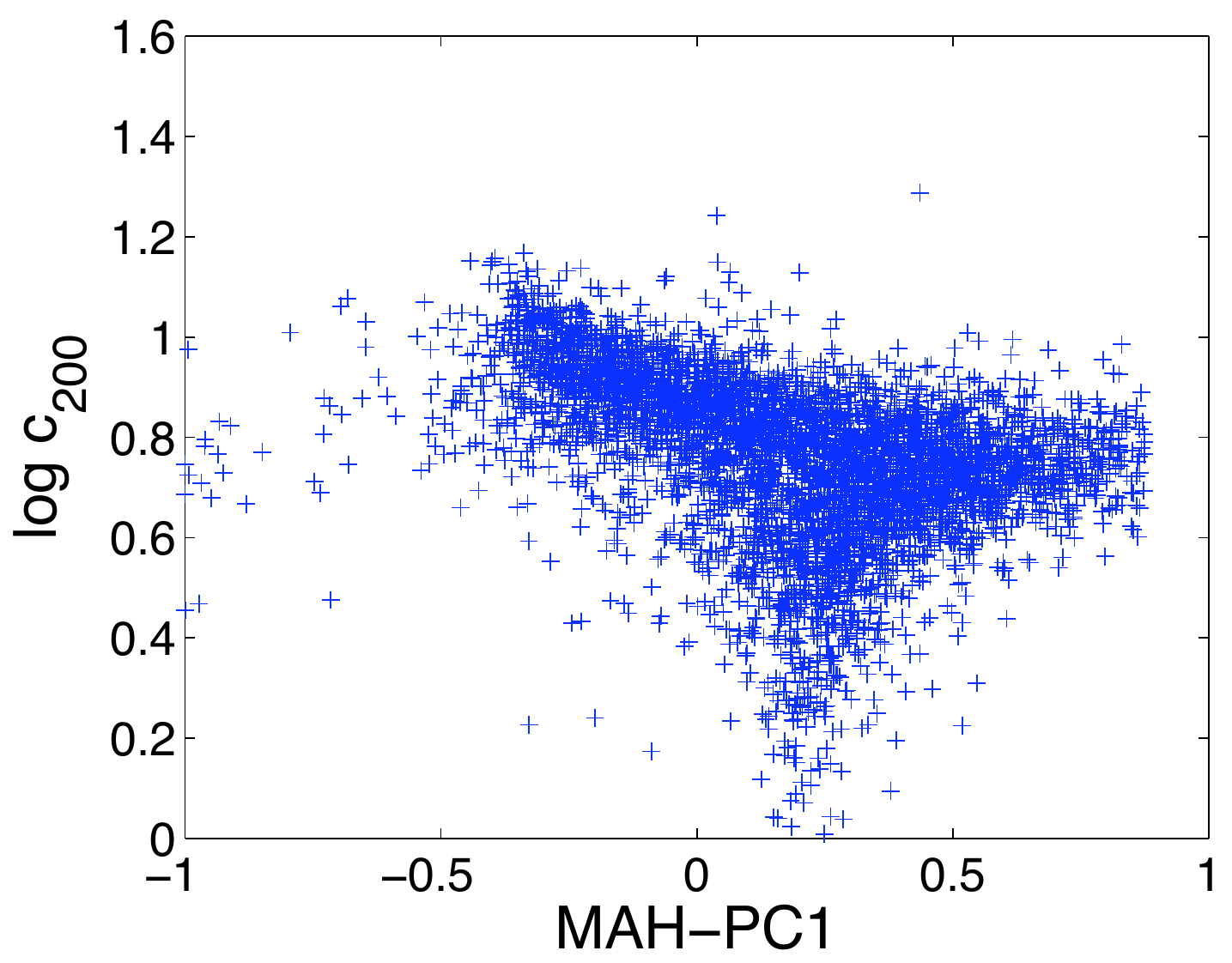}
 \label{subfig:mah_pcproj1_c}
}
\subfigure{
 \includegraphics[width=0.22\textwidth]{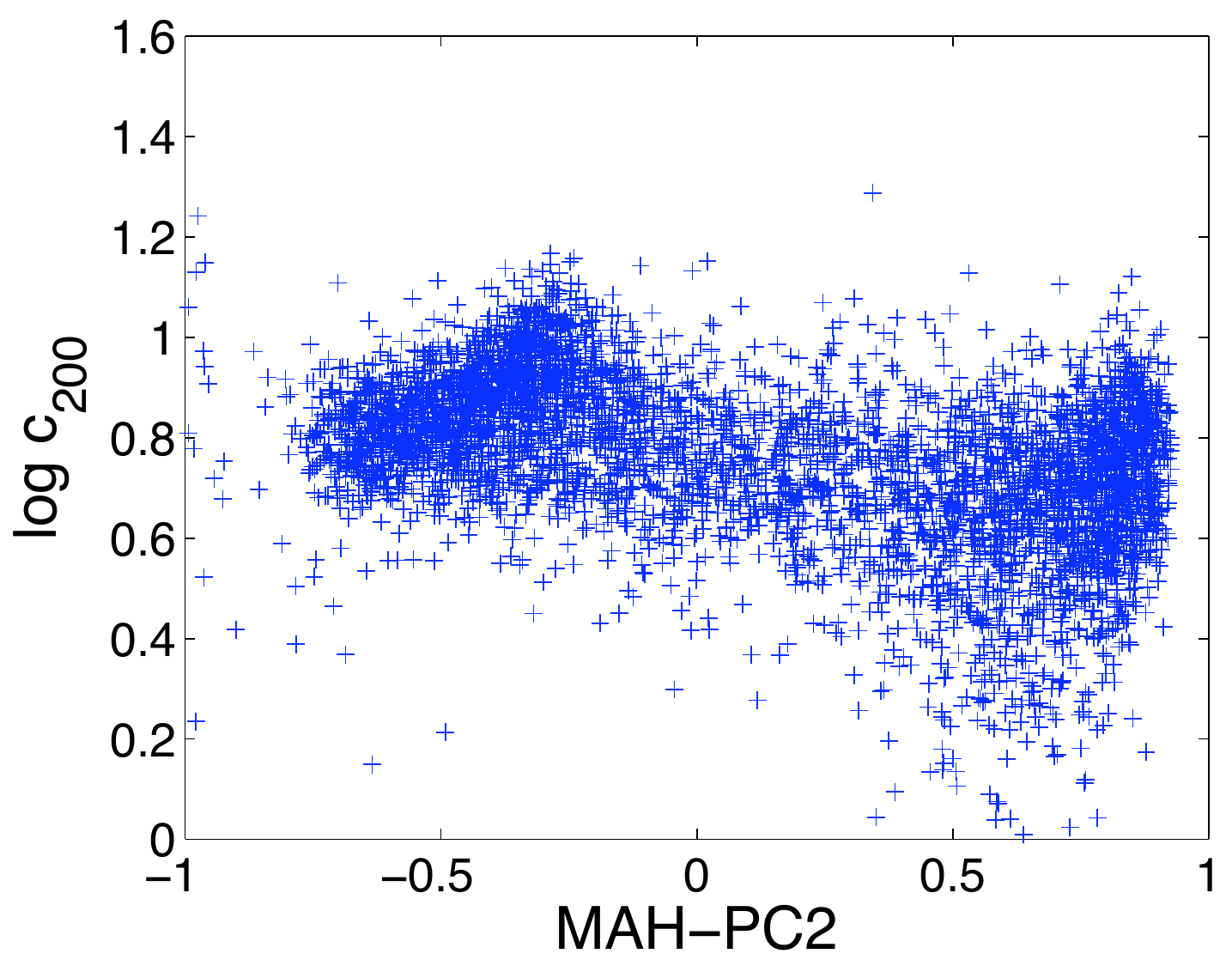}
 \label{subfig:mah_pcproj2_c}
}
\subfigure{
 \includegraphics[width=0.22\textwidth]{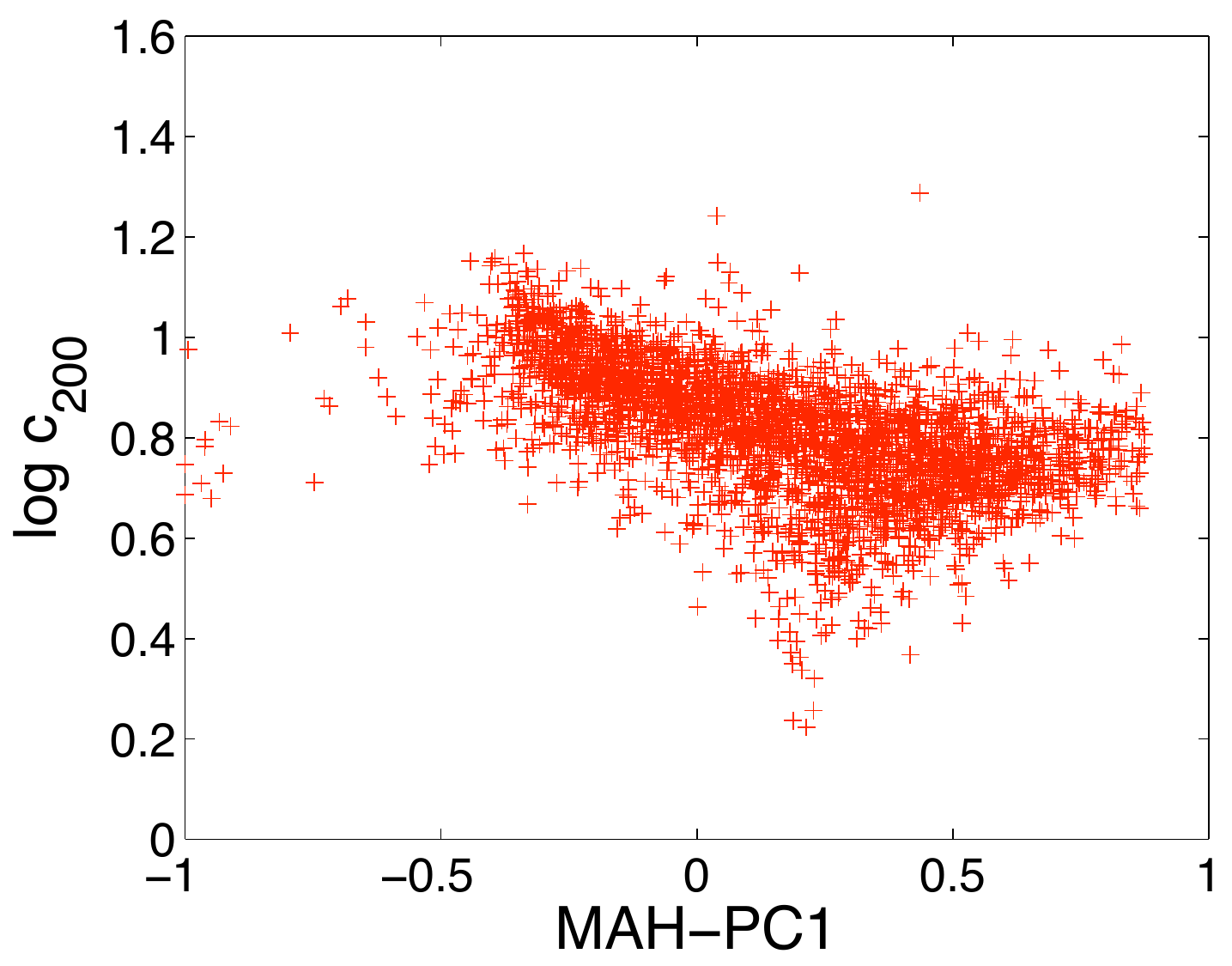}
 \label{subfig:mah_pcproj1_c_relaxed}
}
\subfigure{
 \includegraphics[width=0.22\textwidth]{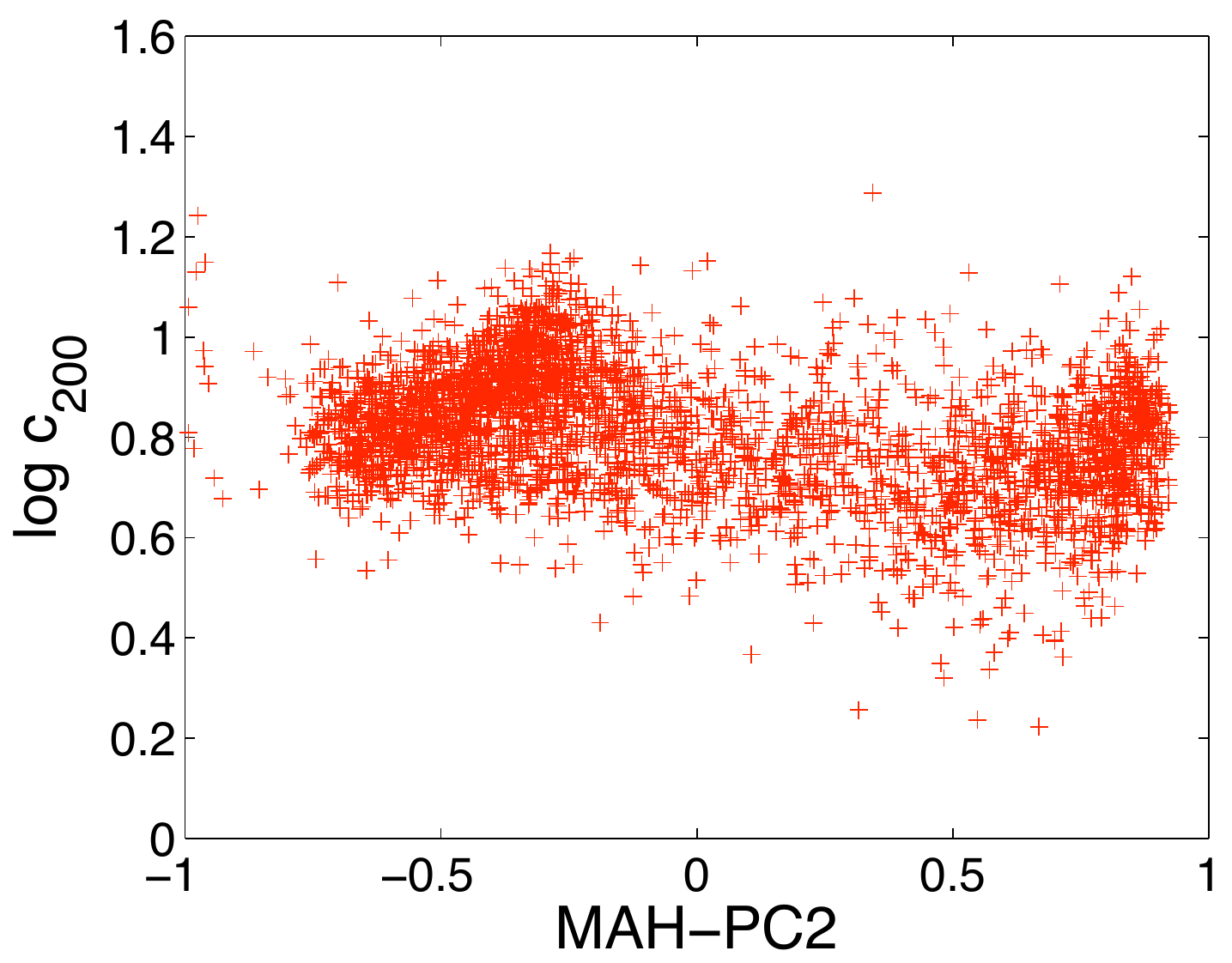}
 \label{subfig:mah_pcproj2_c_relaxed}
}
\subfigure{
 \includegraphics[width=0.22\textwidth]{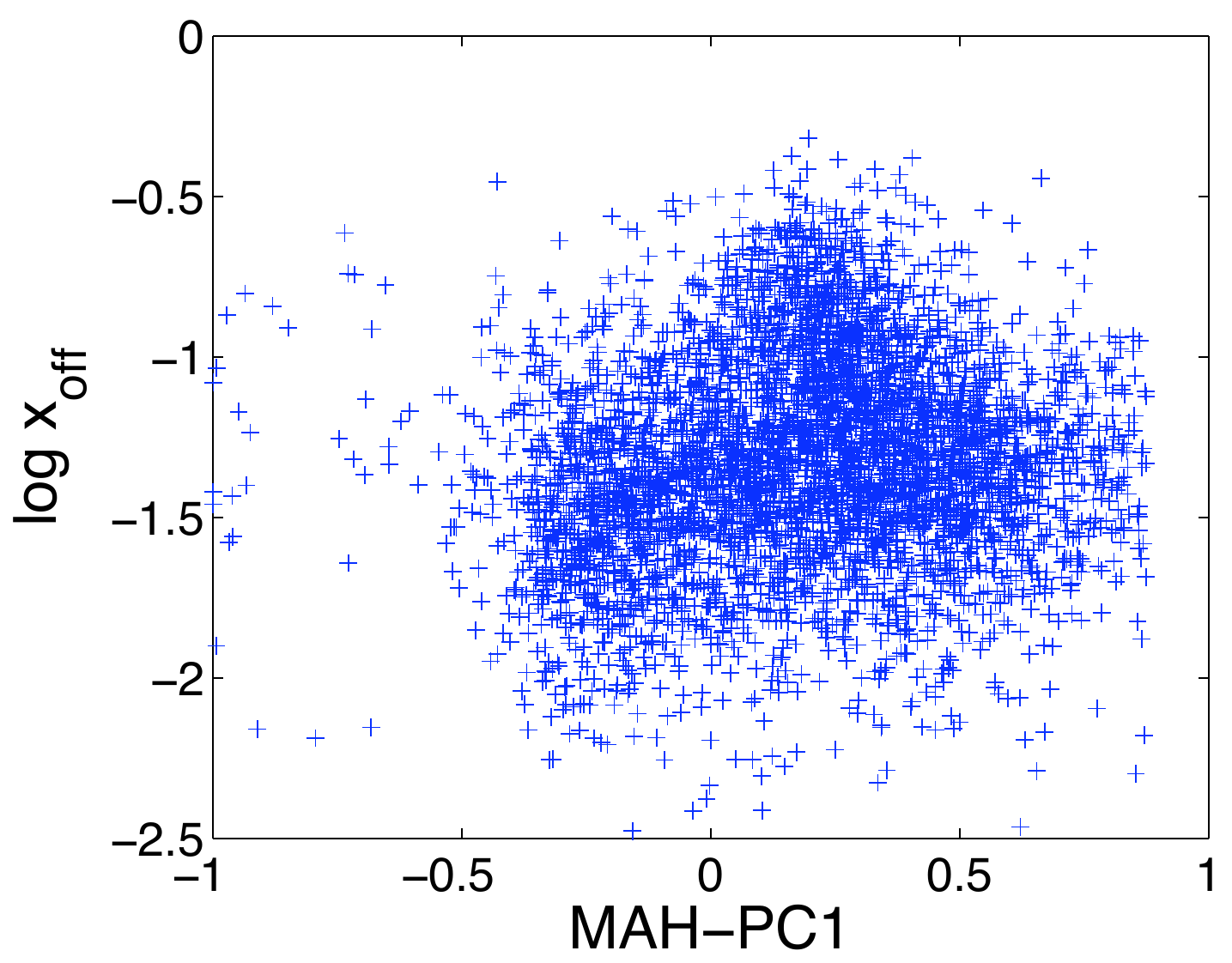}
 \label{subfig:mah_pcproj1_E}
}
\subfigure{
 \includegraphics[width=0.22\textwidth]{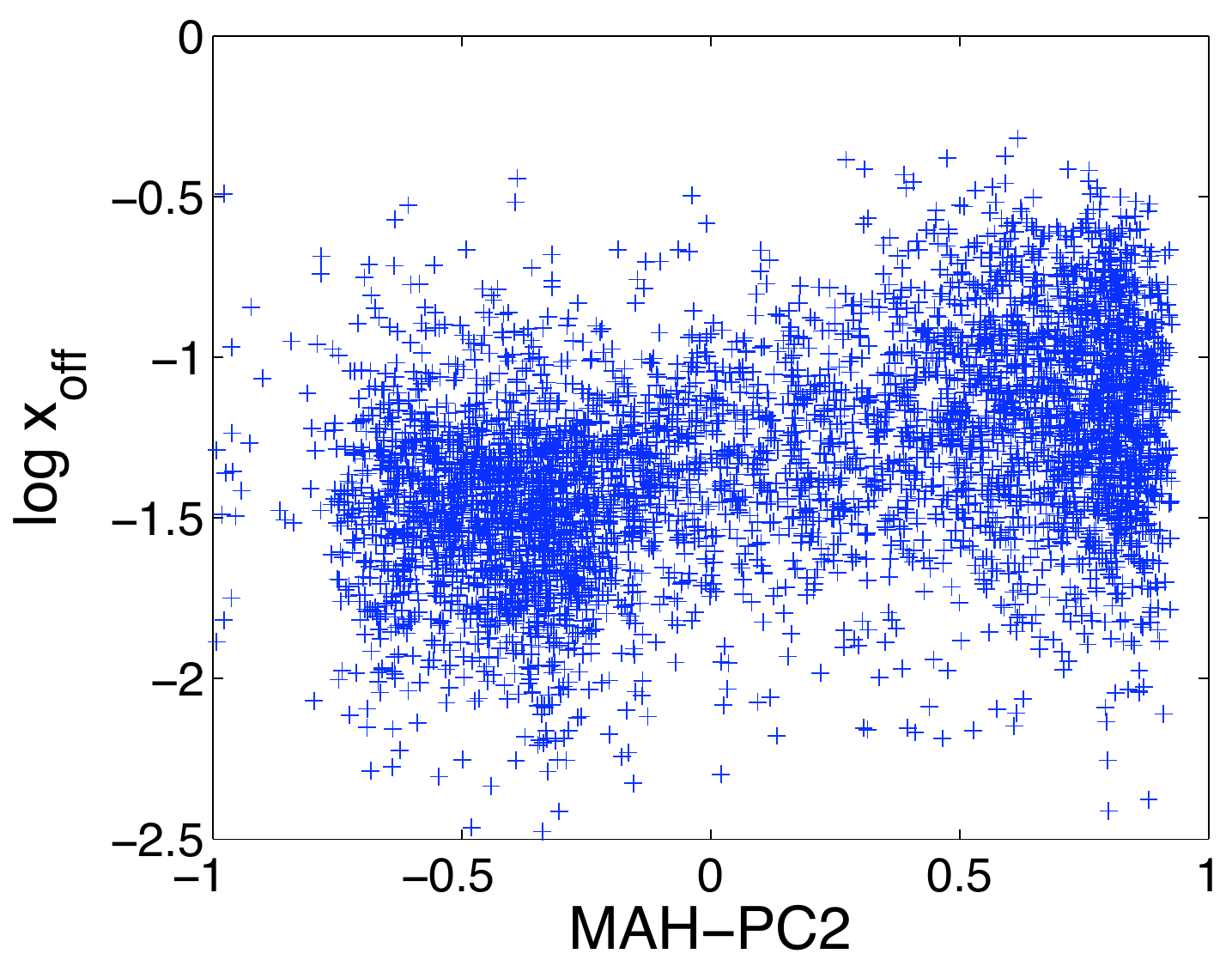}
 \label{subfig:mah_pcproj2_E}
}
\subfigure{
 \includegraphics[width=0.22\textwidth]{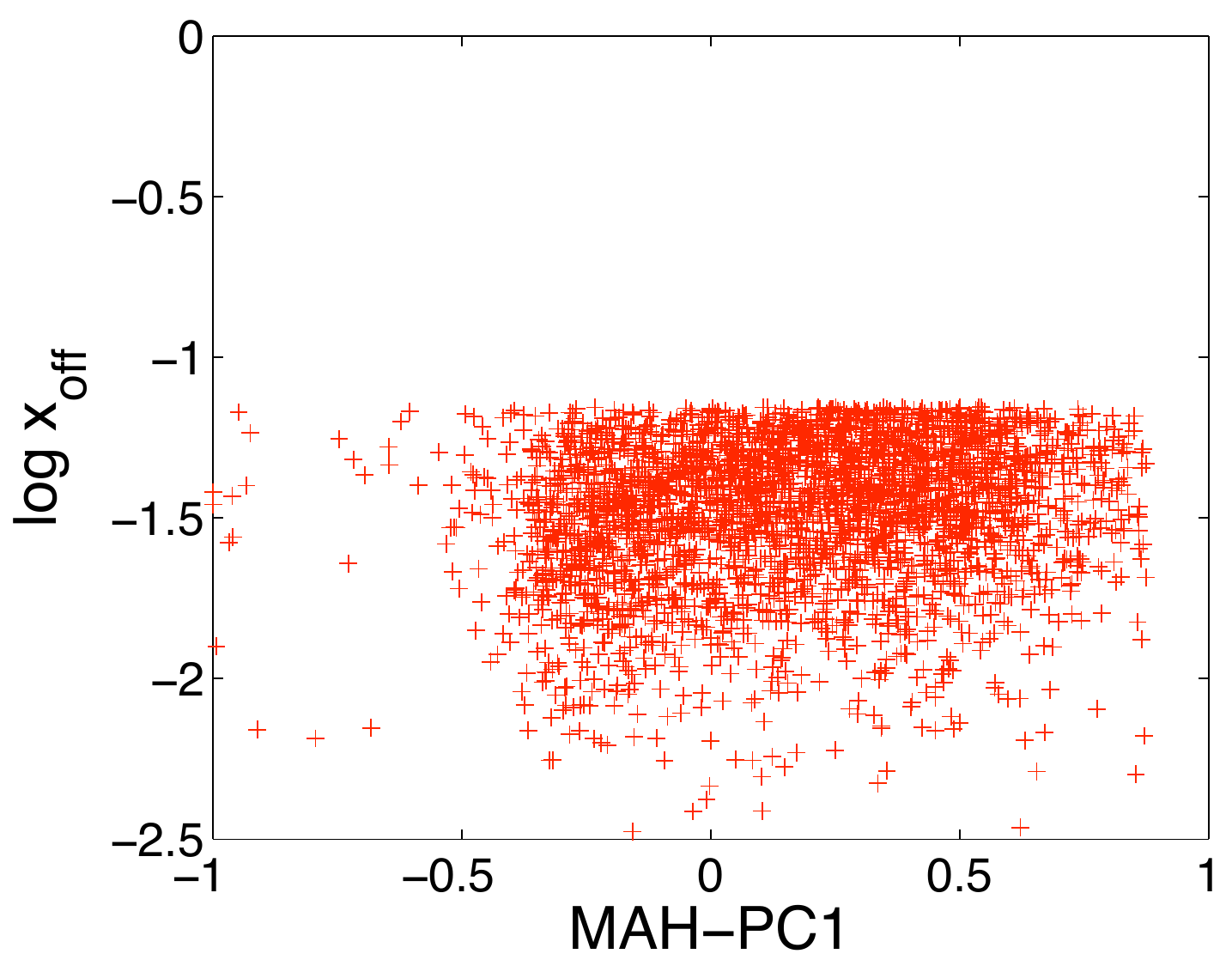}
 \label{subfig:mah_pcproj1_E_relaxed}
}
\subfigure{
 \includegraphics[width=0.22\textwidth]{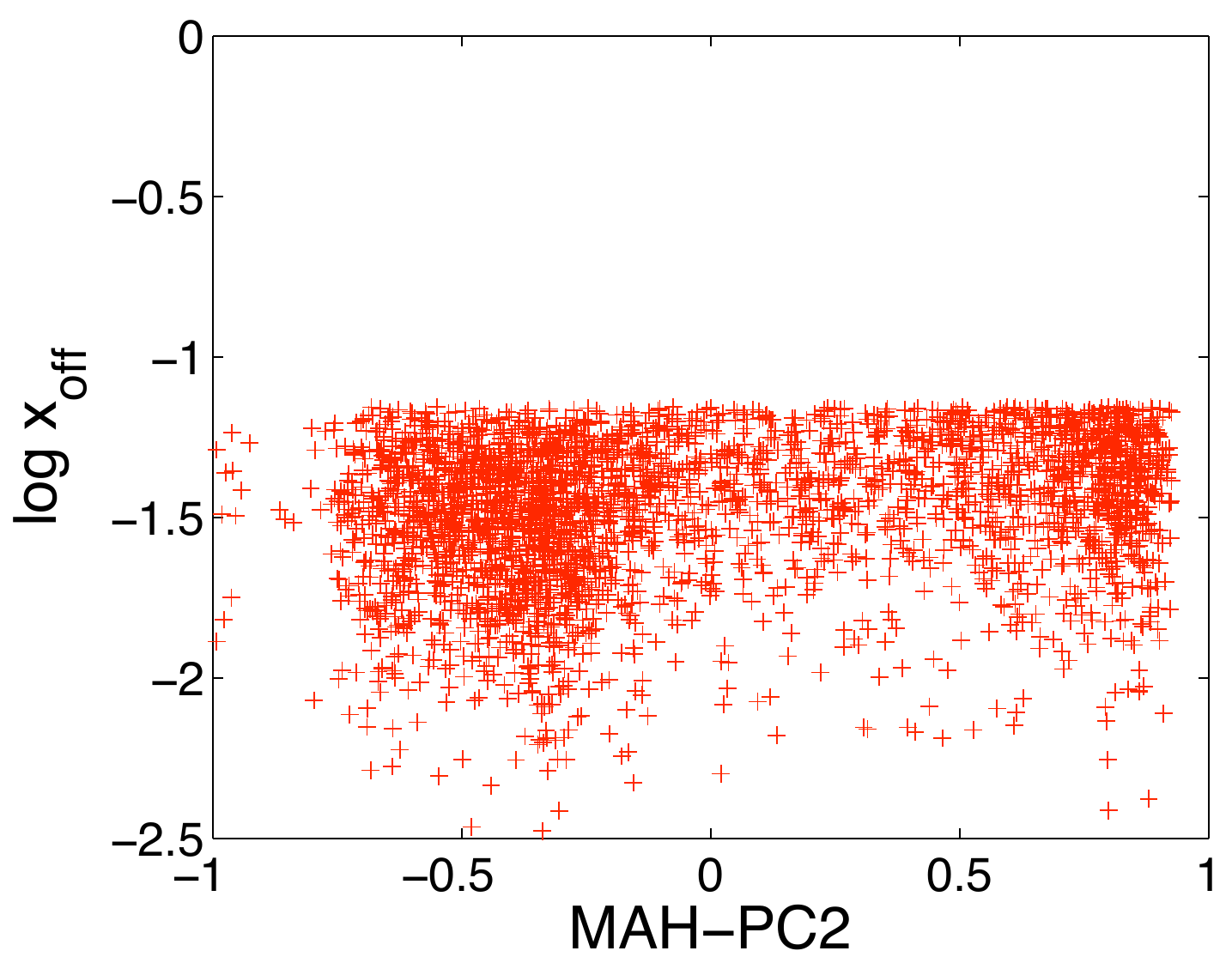}
 \label{subfig:mah_pcproj2_E_relaxed}
}
\caption{Structural parameters versus MAH-PCs for the full sample (left-hand panels) and relaxed subsample (right-hand panels).}
\end{center}
\end{figure*}
\end{center}

\section{Application to Observations}
\label{sec:application}

\subsection{What Do Structural Properties Tell Us?}

Clearly the formation history of dark matter halos, summarized by their MAH, relates to their final structure at a statistical level. Observationally, structure can be measured directly,  whereas history cannot be. Thus a practical application of our results is to infer an unobservable quantity from observable ones. The first question is, given a set of measured structural properties, what to they tell us about the MAH?

\begin{figure}[t]
\centering
\includegraphics[scale=0.60]{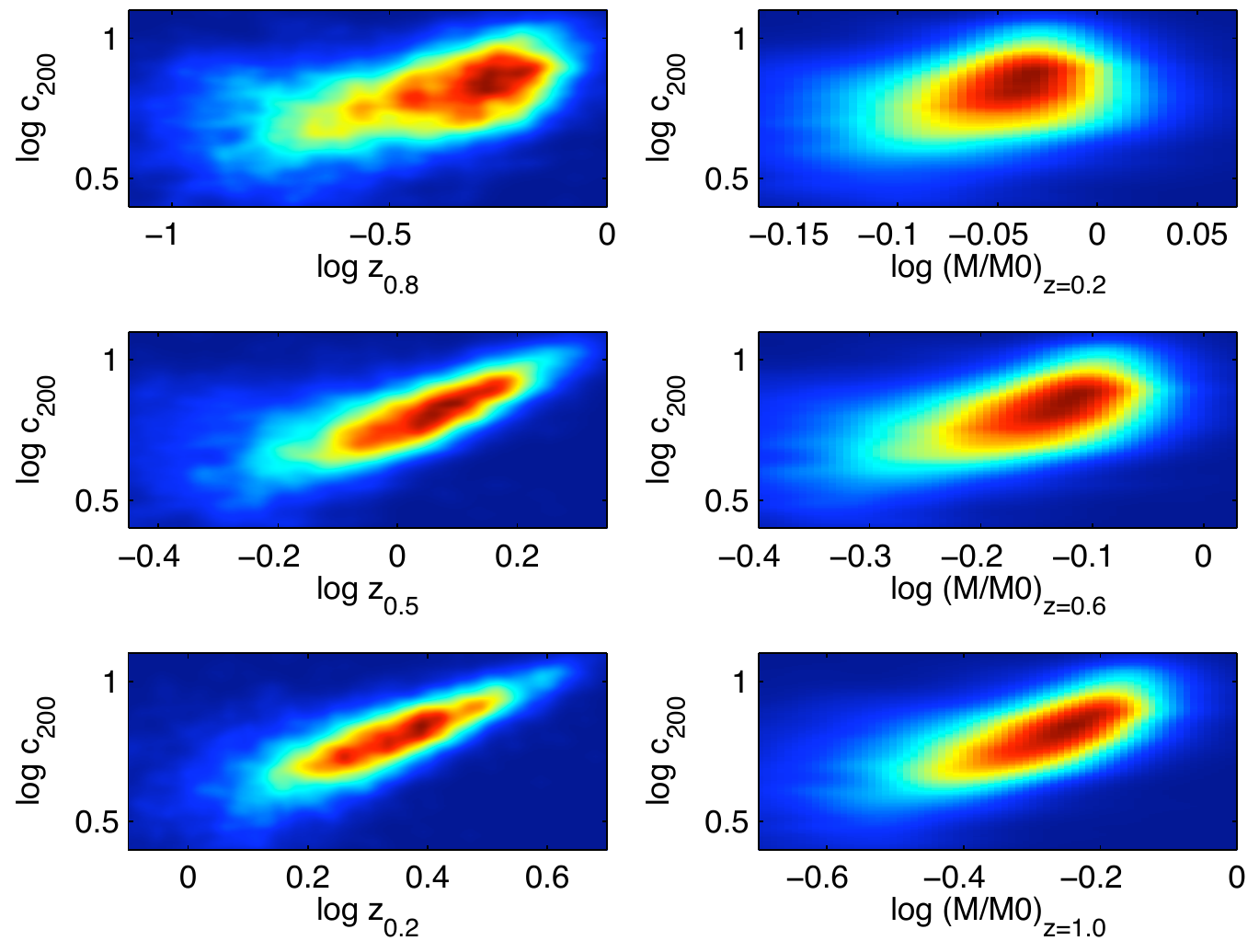}
\caption{Concentration versus age indicators for the full sample.}
\label{fig:correlation_existence}
\end{figure}

From the analysis in Section \ref{subsec:principal_component_analysis}, the parameter most closely and simply related to formation history is the concentration. But which aspect of the formation history does concentration trace most sensitively? Fig. \ref{fig:correlation_existence} shows the correlation between concentration and a number of the age indicators introduced previously. These are $z_x$, the redshift by which the main progenitor of a halo had a fraction $x$ of the final halo mass at $z=0$, for $x = 0.2, 0.5, 0.8$, and $(M/M_0)_z$, the fraction of its final mass a halo had reached by redshift $z$. They represent the values obtained by intersecting the MAH $\mathcal{M}(z)$ with a series of horizontal or vertical lines respectively. Of all the age indicators, $z_{0.2}$ shows the tightest correlation; generally the correlation with $z_x$ is tighter than with $(M/M_0)_z$. Thus concentration is most closely related to the early part of a halo's formation history, when the dense central core of the halo is established.

\begin{figure}[t]
\centering
\includegraphics[scale=0.60]{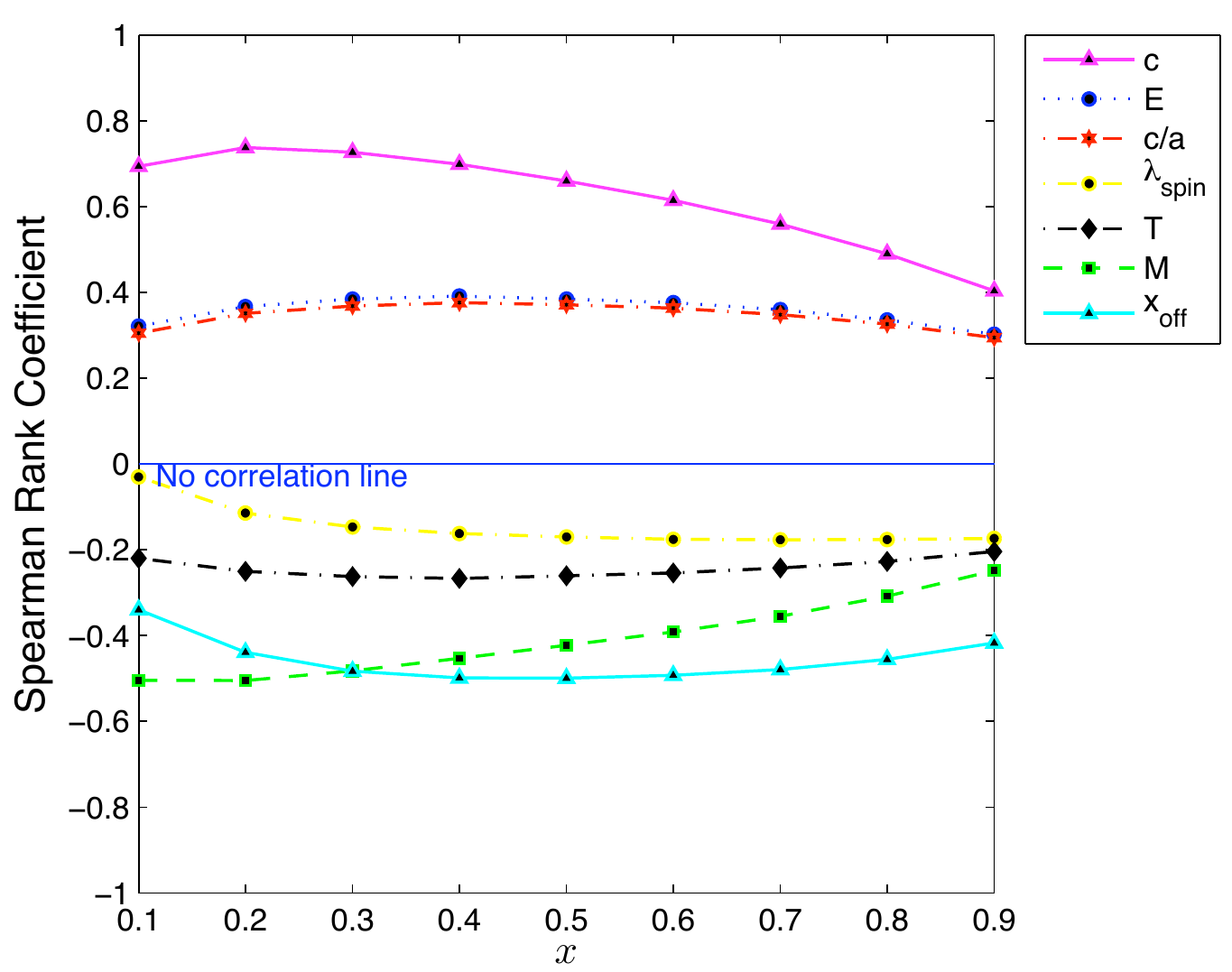}
\caption{Strength of the correlation between structural properties and the formation redshift $z_x$, the redshift by which the halo had built up a fraction $x$ of its final mass at $z = 0$, as a function of $x$.}
\label{fig:spearman_correlation_x}
\end{figure}

We can generalize the specific values of $x$ considered in Fig. \ref{fig:correlation_existence} to a continuous range of values, and test for correlations with other structural parameters. Fig.~\ref{fig:spearman_correlation_x} shows how the strength of the correlation between a given structural parameter and $z_x$ varies as a function of $x$. We use the Spearman rank coefficient as our measure of correlation. Concentration is most strongly correlated with $z_{0.2}$, although the decrease in correlation at $x=0.1$ may be an artifact of our MAH selection and or limited resolution at early times. Elongation $E$ and sphericity $c/a$ are relatively insensitive to $x$, with correlation peaking slightly in the range $x = $0.3--0.5. Relaxedness $x_{\rm off}$ shows a similar pattern of anti-correlation, with a broad peak around $x = $0.4--0.6. Spin and triaxiality are more weakly anticorrelated, with the significance peaking at intermediate or high $x$. We also see that mass $M$ is anti-correlated with $z_x$, particularly for small values of $x$. This conveys the fact that halos with lower values of $z_{0.2}$ will be more massive on average, as expected from the conventional picture of hierarchical structure formation.

\subsection{Splitting Halo Samples by Structural Properties}

\begin{center}
\begin{figure*}
\begin{center}
\subfigure[Age distribution of halos with a split by present-day concentration at $c^{\text{split}}_{200} = \langle c \rangle$ = 6.]{
 \includegraphics[scale=0.5]{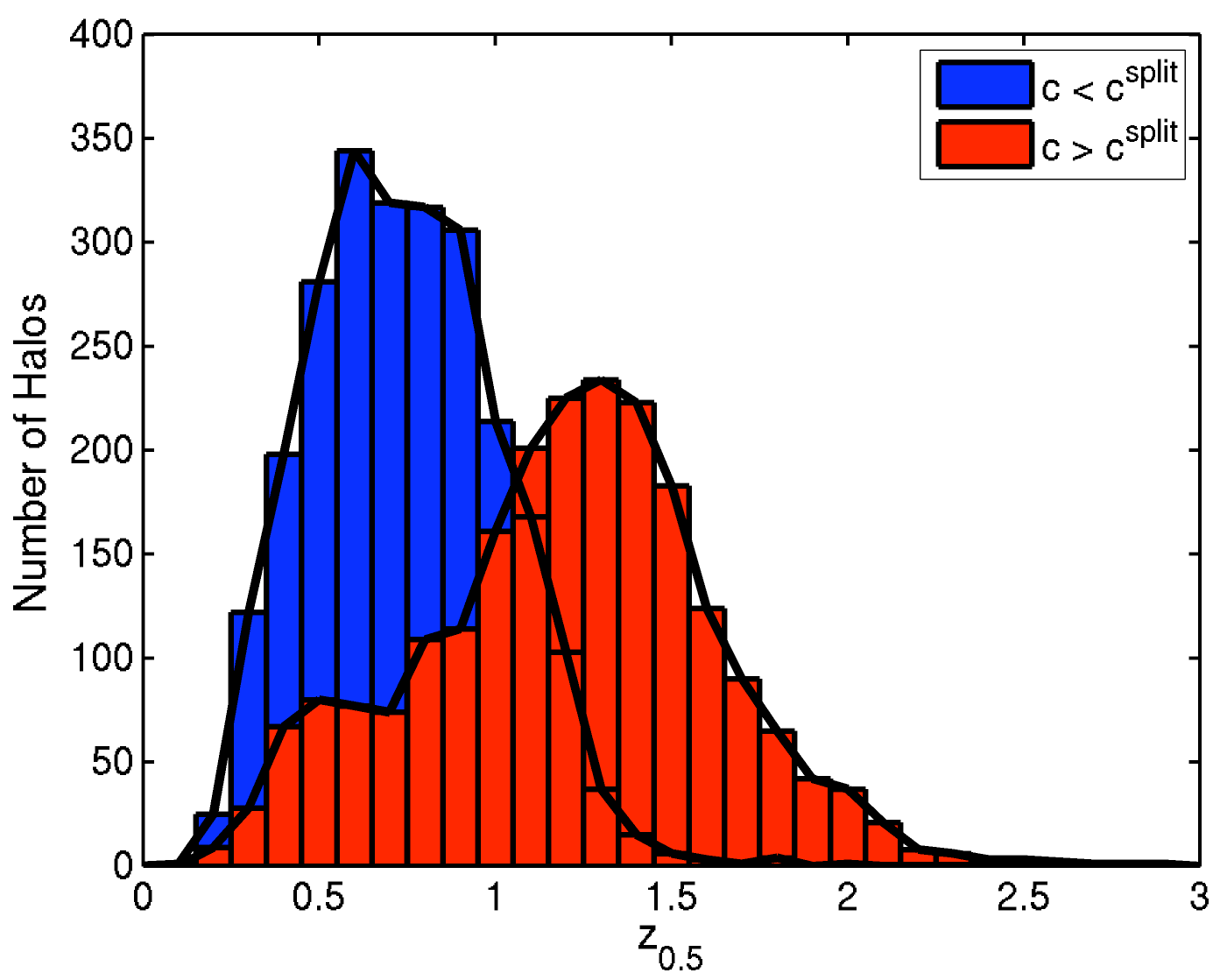}
}
\subfigure[Age distribution of halos with a split by present-day elongation at $E^{\text{split}} = \langle E \rangle$ = 0.4.]{
 \includegraphics[scale=0.5]{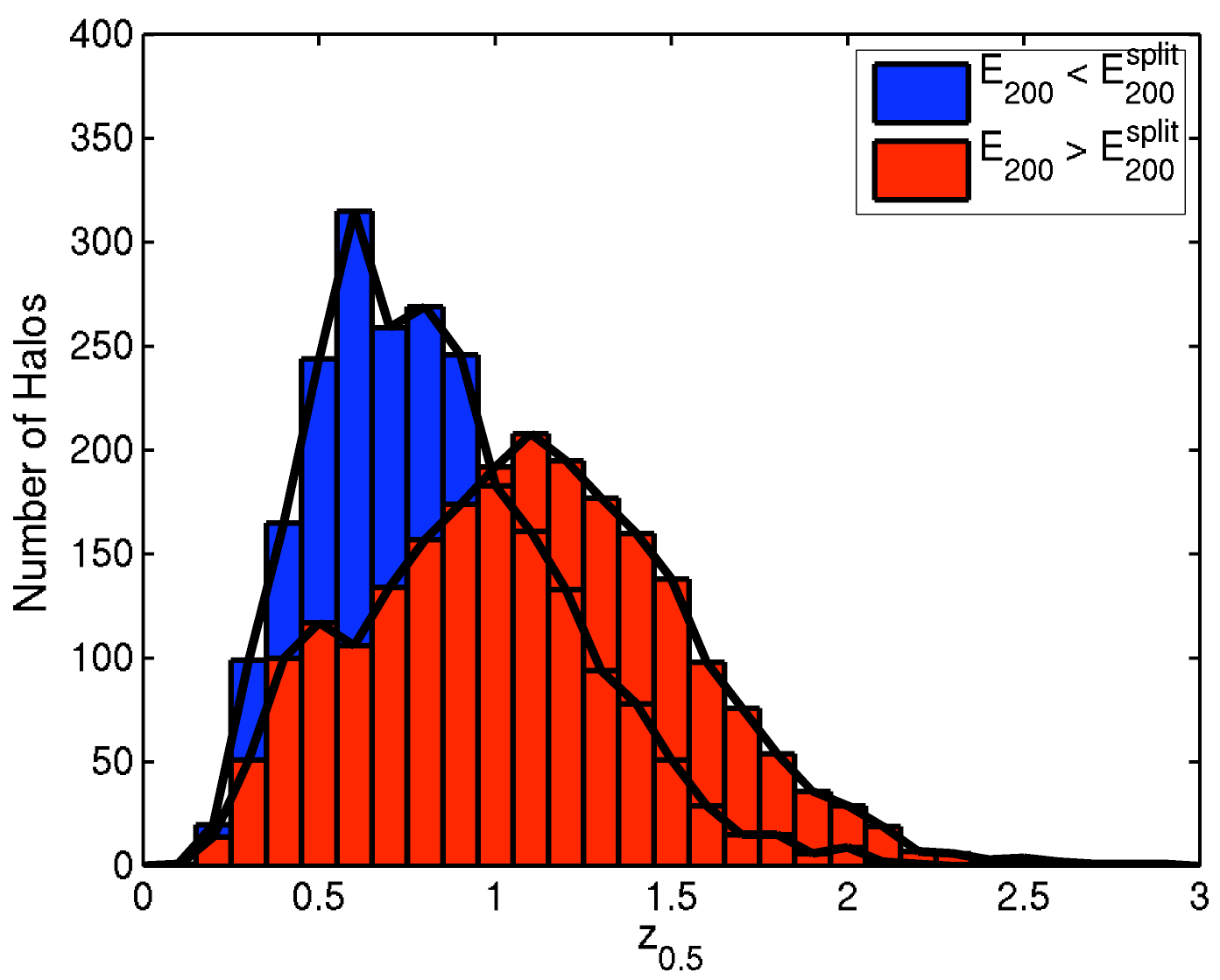}
}
\subfigure[Age distribution of halos with a split by present-day sphericity at $c/a^{\text{split}} = \langle c/a \rangle$ = 0.5.]{
 \includegraphics[scale=0.5]{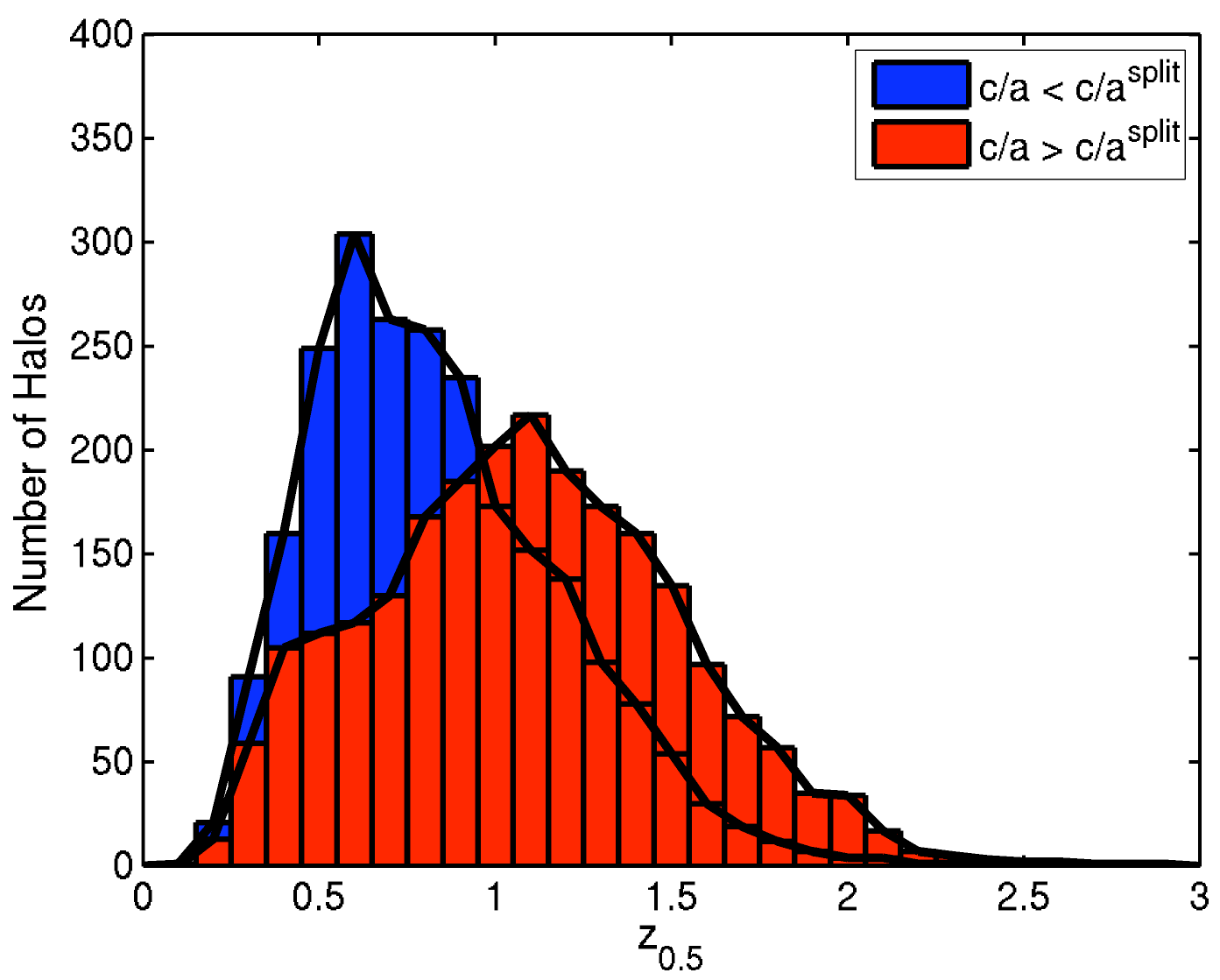}
}
\subfigure[Age distribution for halos with a split by present-day relaxednesses at $x^{\text{split}}_{\t{off}} = \langle x_{\t{off}} \rangle$ = 0.09.]{
 \includegraphics[scale=0.5]{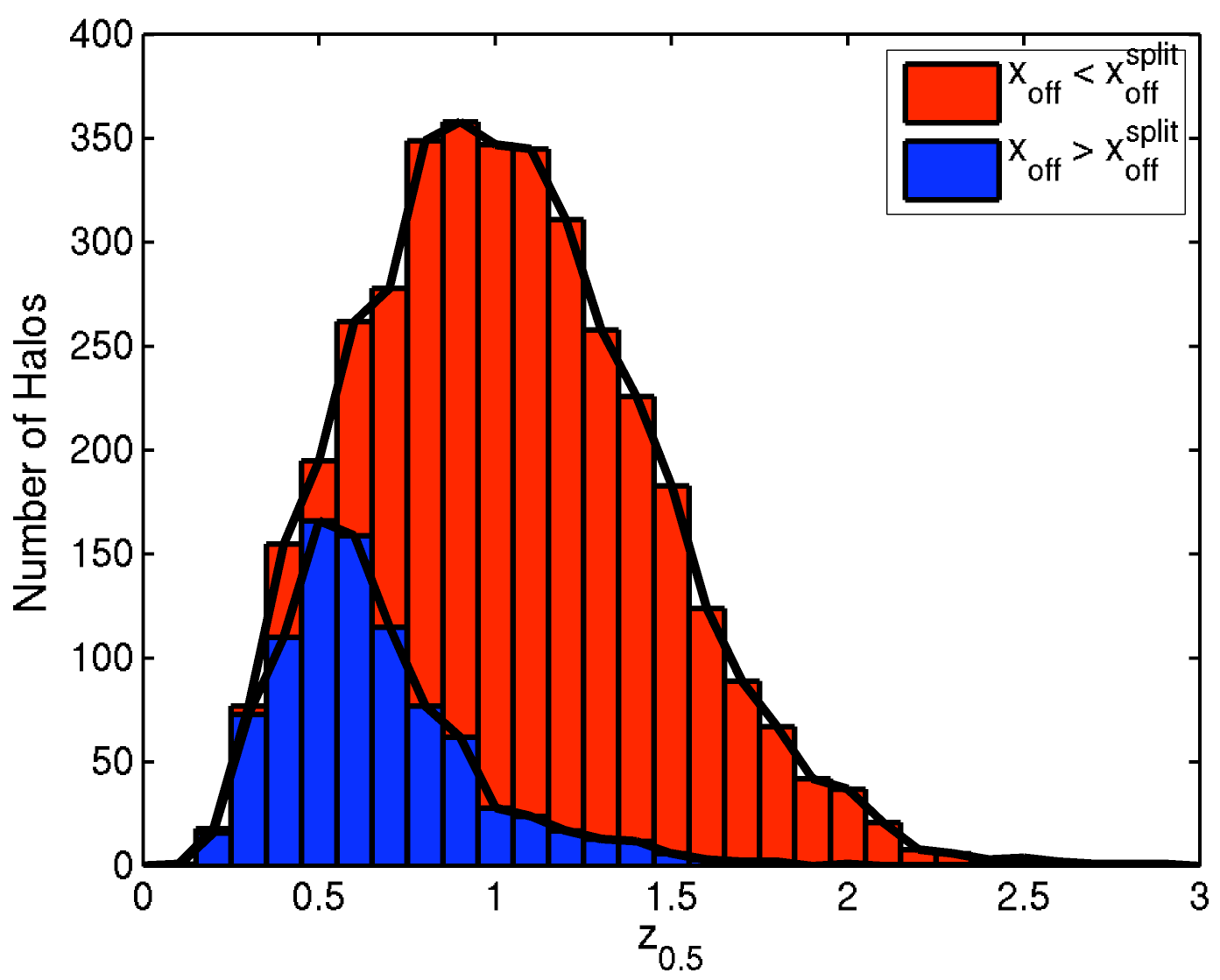}
}
\caption{Distributions of the formation redshift $z_{0.5}$ in halo samples split by present-day structural properties}
\label{fig:split_zf}
\end{center}
\end{figure*}
\end{center}

Given the scatter in all the correlations seen so far between structure and formation history, the measurement of structural parameters in any single system will give only a rough indication of its past history. A more interesting possibility is to average over sets of objects. As mentioned in Section \ref{sec:introduction}, structural properties such as shape and concentration have now been measured for hundreds of galaxy cluster halos, and may soon be measured reliably for galaxy halos in forthcoming weak lensing surveys. In the case of the most massive clusters, it may even be possible to constrain the full 3-dimensional shape by combining lensing with other observations \citep[e.g.][]{Morandi2011-xraylensing}. In Fig.~\ref{fig:split_zf}, 
we show how, by splitting a sample of halos into subsamples based on structural properties at $z=0$, we can construct subsamples with systematically different formation histories, as indicated by the distribution of $z_{0.5}$ values. In the four panels we consider splits based on concentration, elongation, sphericity and relaxedness. Consistent with our earlier results, we find that a split based on concentration produces the largest offset in the mean value of $z_{0.5}$ between the
two subsamples, while splits based on shape produce similar but smaller offsets. 
(These offsets will also be further reduced in most practical tests, since for most systems we will observe only projected shape, not true 3-D shape.)
Relaxedness (which could be determined from detailed mapping of the mass distribution, e.g.~in the X-ray) does not produce much shift in $z_{0.5}$ if we split 
the sample evenly, but it does select out systematically younger halos if we cut out the minority of very unrelaxed systems. The cut illustrated in Fig.~\ref{fig:split_zf}
is close to the one we have use throughout the paper to separate relaxed and unrelaxed systems. A similar selection by relaxedness has already been 
applied a pilot sample of 10 clusters by \cite{smithtaylor} in order to study the connection between X-ray morphology, offsets from X-ray scaling relations, 
and formation history, but larger samples are needed to reach definitive conclusions.

\begin{center}
\begin{figure*}
\begin{center}
\subfigure[Splitting above and below $E^{\text{split}} = \langle E_{z=0.5} \rangle$ = 0.4.]{
 \includegraphics[scale=0.5]{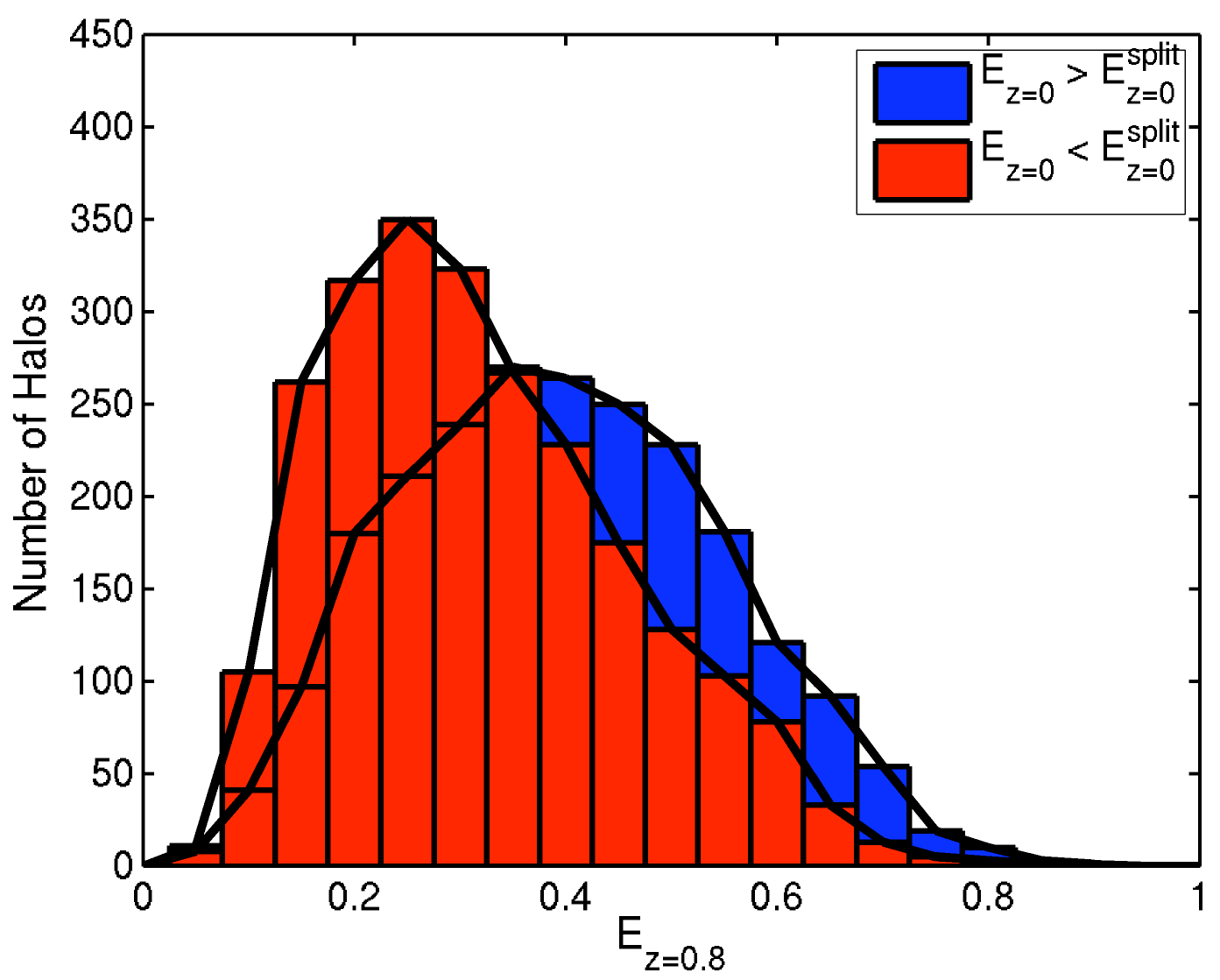}
}
\subfigure[Splitting above and below $c/a^{\text{split}}_{z=0.5} = \langle c/a_{z=0.5} \rangle$ = 0.5.]{
 \includegraphics[scale=0.5]{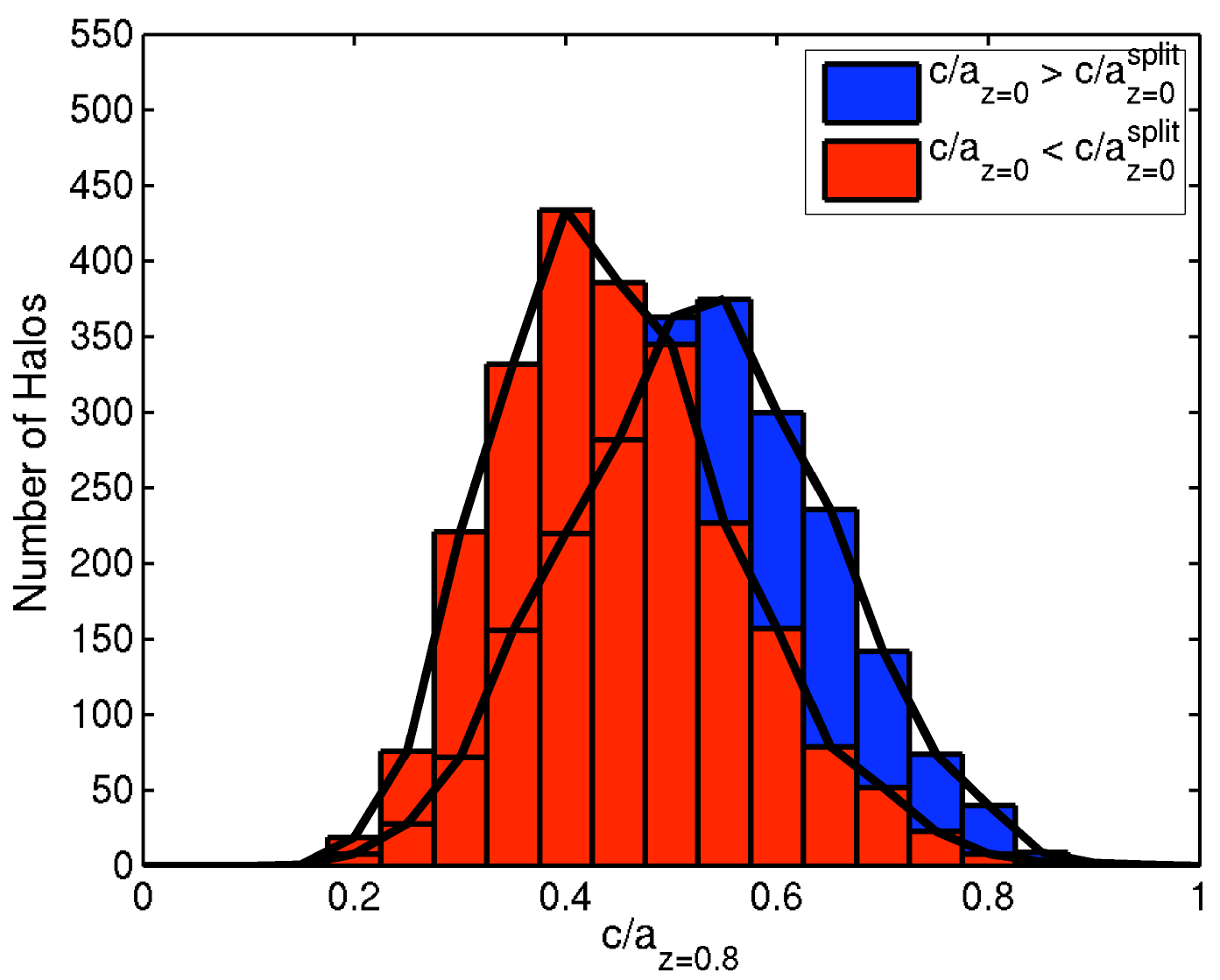}
}
\caption{Distributions of the shape of the progenitor region from which a halo formed,  for halo samples split by their shape at $z=0$.}
\label{fig:split_shape}
\end{center}
\end{figure*}
\end{center}

Splits based on present-day samples can also tell us other things about the past history of a population of halos. In Fig.~\ref{fig:split_shape}
we show how cuts based on elongation or sphericity at $z=0$ select halos formed from material with a systematically different spatial distribution at $z = 0.5$.
This might eventually provide interesting tests of non-Gaussianity in structure formation, or test the dependence of galaxy formation on the detailed environment around a group or cluster. 

\section{Summary}
\label{sec:summary}

The recent studies of S11 and J11 have clarified how the structural properties of halos are inter-related, and how they relate to a few specific 
measures of the formation history. Here we have extended this approach to study the connection between structure and formation history in more detail. 
We first examine the intrinsic variation in halo formation histories, as summarized by their MAHs $\mathcal{M}(z)$. Applying principal component analysis
to the individual steps of the MAHs themselves, we can decompose the range of formation history into an mean MAH and a main set of variations about 
the mean. The mean MAH and variations away from it along the first principal axis {\bf MAH-PC}$_1$ are well-fit by the 2-parameter function suggested by 
\citet{McBride2009-MAH}, although the two parameters $\beta$ and $\gamma$ are not a natural parameterization of this variation. 
Since {\bf MAH-PC}$_1$ is the strongest principal component, and since the variations along this axis correspond to MAHs that 
reach a given fraction of the final mass at systematically earlier or later times, it is natural to consider this the best (MAH-based) definition of halo age.

A second component, {\bf MAH-PC}$_2$, is also relatively important in the sense that it accounts for 17\%\ of the total variance in the MAHs. This component 
corresponds to acceleration or deceleration in the accretion rate at late times. It is not particularly well fit by the McBride formula. Third and higher 
PCs correspond roughly to terms in a Fourier-like decomposition of the MAH, with an additional oscillation about the mean
in each successive PC, but we have not studied them in detail since they are less 
important sources of scatter individually. Collectively, however, they account for almost 30\%\ of the total variance, showing that individual MAHs 
are not particularly well fit by smooth curves. This is presumably due to the large stochastic jumps produced by major mergers.

Relating formation history and structural parameters, we recover the trends seen in the previous studies. Concentration, in particular, correlates 
strongly with age indicators. Studying the relationship between concentration and {\bf MAH-PC}$_1$, however, we find that it is bimodal. Relaxed systems
have a fairly tight correlation between age (as expressed by {\bf MAH-PC}$_1$) and concentration, but in disturbed, low-concentration systems this correlation 
breaks down. Thus concentration provides an effective measure of age, but only in relaxed systems. Testing the strength of the correlation against various 
age-related parameters, we find it is tightest for the formation redshift $z_{0.2}$, the redshift by which a halo had built up 20\%\ of its final mass in 
its most massive progenitor, or equivalently the redshift where $\mathcal{M}(z) = 0.2$. Thus concentration is an indication of early merger history. 

Testing for correlations with other parameters, we find that shape and relaxedness are better tracers of the more recent formation history of 
a halo. Shape parameters such as elongation or sphericity are reasonably well correlated with the shape of the mass distribution from which the 
halo formed, but only going back to low redshifts, that is to say only considering a halo's recent past. Similarly, irregularities in the density profile 
seem to correlate more with the last half of a halo's growth history.

These theoretical results have several possible applications to real systems. In the case of individual systems, concentration can give us a broad 
indication of early history, while relaxedness can give us an indication of later history. Perhaps more interesting, however, are the implications for 
large samples of halos. Splitting cluster catalogs by concentration, shape, or relaxation should select sub-samples with systematically different 
formation histories. Comparing these to predicted MAH distributions could offer new cosmological tests, updating an early idea of testing cosmology 
by measuring the ages of clusters, as indicated by substructure \citep{Richstone}. On galaxy scales, if weak lensing surveys can reach the precision where halo shape is measured 
routinely, this may allow a direct empirical test of how the MAH influences galaxy formation.

\acknowledgments
This work was supported by a Discovery Grant to JT and an Undergraduate Summer Research Award to AW, both from the Natural Sciences and Engineering Research Council (NSERC) of Canada. The simulations were performed using the facilities of the Shared Hierarchical Academic Research Computing Network (SHARCNET -- www.sharcnet.ca) and Compute/Calcul Canada. We thank SHARCNET staff for their technical support throughout the project.

\bibliography{PCAreferences}

\appendix

\section{Correlation Tables for Halo Properties}
\label{app:full_correlation_table}

\begin{figure}[b]
\centering
\includegraphics[scale=0.7]{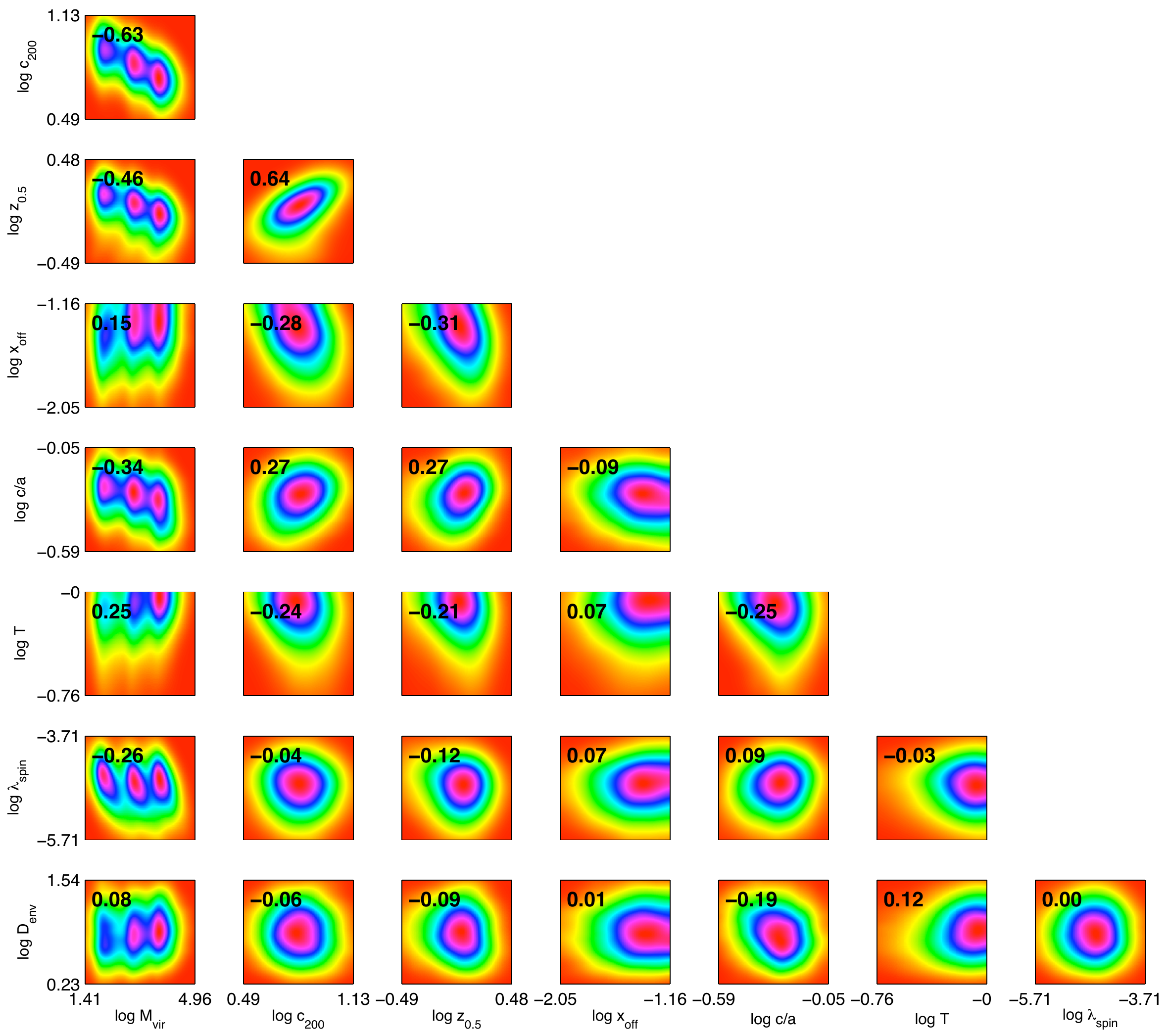}
\caption{Spearman correlation table for the 7 parameters also used in \cite{JeesonDaniel2011-PCAcorrelations}.}
\label{fig:correlation_full_table_compare_jeeson}
\end{figure}

\begin{figure}
\centering
\includegraphics[scale=0.7]{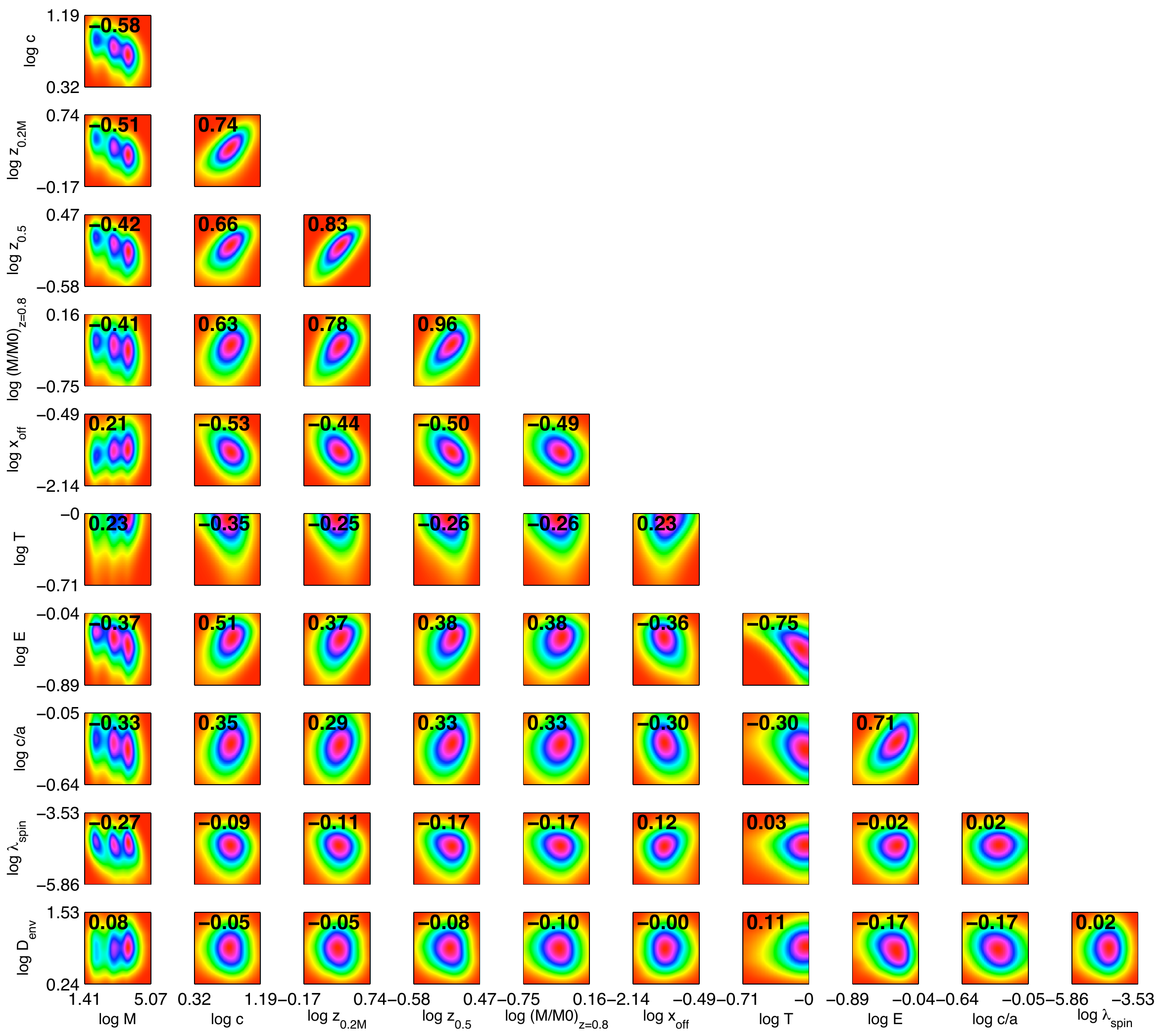}
\caption{A full Spearman correlation table of the parameters discussed in Section \ref{subsec:chosen_halo_properties}. 
The values indicated on the axes are the base-10 log of the parameter in question.}
\label{fig:correlation_full_table}
\end{figure}

Correlation plots and Spearman coefficients using the subset of 7 parameters also used by J11 is shown 
in Fig. \ref{fig:correlation_full_table_compare_jeeson}, to aid comparison with this previous work.
The full correlation analysis of the halo structural properties is shown in Fig. \ref{fig:correlation_full_table}.

\end{document}

%% file: pca_table2.tex
\begin{table}[H]
\caption{Principal components of the full sample}
\centering
\begin{tabular}{| c | c | c | c | c |}
\hline
Principal axis & \bf{S-PC}$_1$ & \bf{S-PC}$_2$ & \bf{S-PC}$_3$ & \bf{S-PC}$_4$ \\
\hline\hline
$\lambda_i/\lambda_{1}$ 	& { 1.00 } &{ 0.34 } & { 0.27 } & { 0.22 } \\ \hline
Contribution       	& { 40\% } &{ 14\% } & { 11\% } & { 9\% } \\ \hline \hline
log $M_{\text{vir}}$ & { 0.29 } &{ 0.11 } & {\bf 0.53 } & { -0.08 } \\ \hline
log $c_{200}$        	& {\bf -0.39 } &{ 0.04 } & { -0.04 } & { -0.13 } \\ \hline
log $z_{0.2}$        	& {\bf -0.39 } &{ 0.29 } & { -0.10 } & { 0.06 } \\ \hline
log $z_{0.5}$        	& {\bf -0.39 } &{\bf 0.32 } & { 0.01 } & { 0.12 } \\ \hline
log $(M/M_0)_{0.5}$  	& {\bf -0.37 } &{\bf 0.33 } & { 0.01 } & { 0.16 } \\ \hline
log $x_{\text{off}}$ 	& {\bf 0.30 } &{ -0.05 } & { -0.10 } & { 0.26 } \\ \hline
log $T$              	& { 0.22 } &{\bf 0.42 } & { -0.21 } & { 0.21 } \\ \hline
log $E$              	& {\bf -0.34 } &{\bf -0.46 } & { 0.21 } & { -0.12 } \\ \hline
log $c/a$            	& { -0.27 } &{\bf -0.41 } & { 0.08 } & { 0.05 } \\ \hline
log $\lambda$        	& { 0.03 } &{ -0.28 } & {\bf -0.76 } & { 0.01 } \\ \hline
log $D_{1,0.1}$      	& { 0.07 } &{ 0.23 } & { -0.15 } & {\bf -0.90 } \\ \hline
\end{tabular}
\label{table:PCA_results_all}
\end{table}

%% file: pca_table1.tex
\begin{table}[H]
\caption{Principal components of the relaxed sample} 
\centering
\begin{tabular}{| c | c | c | c | c |}
\hline
Principal axis & \bf{S-PC}$_1$ & \bf{S-PC}$_2$ & \bf{S-PC}$_3$ & \bf{S-PC}$_4$ \\
\hline\hline
$\lambda_i/\lambda_{1}$ 	& { 1.00 } &{ 0.43 } & { 0.29 } & { 0.24 } \\ \hline
Contribution       	& { 36\% } &{ 16\% } & { 10\% } & { 9\% } \\ \hline \hline
log $M_{\text{vir}}$ 	& {\bf 0.34 } &{ 0.11 } & {\bf -0.46 } & { -0.08 } \\ \hline
log $c_{200}$        	& {\bf -0.39 } &{ 0.10 } & { 0.19 } & { -0.06 } \\ \hline
log $z_{0.2}$        	& {\bf -0.41 } &{ 0.27 } & { 0.06 } & { 0.02 } \\ \hline
log $z_{0.5}$        	& {\bf -0.40 } &{\bf 0.32 } & { -0.06 } & { 0.08 } \\ \hline
log $(M/M_0)_{0.5}$  	& {\bf -0.36 } &{\bf 0.30 } & { -0.11 } & { 0.17 } \\ \hline
log $x_{\text{off}}$ 	& { 0.17 } &{ -0.16 } & { 0.04 } & {\bf 0.36 } \\ \hline
log $T$              	& { 0.21 } &{\bf 0.36 } & { 0.20 } & {\bf 0.43 } \\ \hline
log $E$              	& {\bf -0.33 } &{\bf -0.49 } & { -0.20 } & { -0.15 } \\ \hline
log $c/a$            	& { -0.28 } &{\bf -0.43 } & { -0.13 } & { 0.08 } \\ \hline
log $\lambda$        	& { 0.01 } &{\bf -0.31 } & {\bf 0.72 } & { 0.19 } \\ \hline
log $D_{1,0.1}$      	& { 0.10 } &{ 0.20 } & {\bf 0.35 } & {\bf -0.76 } \\ \hline
\end{tabular}
\label{table:PCA_results_relaxed}
\end{table}